\documentclass[]{article}
\usepackage{lmodern}
\usepackage{amssymb,amsmath}
\usepackage{ifxetex,ifluatex}
\usepackage{fixltx2e} 
\ifnum 0\ifxetex 1\fi\ifluatex 1\fi=0 
  \usepackage[T1]{fontenc}
  \usepackage[utf8]{inputenc}
\else 
  \ifxetex
    \usepackage{mathspec}
  \else
    \usepackage{fontspec}
  \fi
  \defaultfontfeatures{Ligatures=TeX,Scale=MatchLowercase}
\fi
\IfFileExists{upquote.sty}{\usepackage{upquote}}{}
\IfFileExists{microtype.sty}{%
\usepackage{microtype}
\UseMicrotypeSet[protrusion]{basicmath} 
}{}
\usepackage[margin=1in]{geometry}
\usepackage{hyperref}
\hypersetup{unicode=true,
            pdftitle={The Locally Gaussian Partial Correlation},
            pdfauthor={Håkon Otneim; Dag Tjøstheim},
            pdfborder={0 0 0},
            breaklinks=true}
\usepackage{longtable,booktabs}
\usepackage{graphicx,grffile}
\makeatletter
\def\maxwidth{\ifdim\Gin@nat@width>\linewidth\linewidth\else\Gin@nat@width\fi}
\def\maxheight{\ifdim\Gin@nat@height>\textheight\textheight\else\Gin@nat@height\fi}
\makeatother
\setkeys{Gin}{width=\maxwidth,height=\maxheight,keepaspectratio}
\IfFileExists{parskip.sty}{%
\usepackage{parskip}
}{
\setlength{\parindent}{0pt}
\setlength{\parskip}{6pt plus 2pt minus 1pt}
}
\providecommand{\tightlist}{%
  \setlength{\itemsep}{0pt}\setlength{\parskip}{0pt}}
\ifx\paragraph\undefined\else
\let\oldparagraph\paragraph
\renewcommand{\paragraph}[1]{\oldparagraph{#1}\mbox{}}
\fi
\ifx\subparagraph\undefined\else
\let\oldsubparagraph\subparagraph
\renewcommand{\subparagraph}[1]{\oldsubparagraph{#1}\mbox{}}
\fi

\let\rmarkdownfootnote\footnote%
\def\footnote{\protect\rmarkdownfootnote}

\usepackage{titling}

  \title{The Locally Gaussian Partial Correlation}
    \pretitle{\vspace{\droptitle}\centering\huge}
  \posttitle{\par}
    \author{Håkon Otneim\footnote{Corresponding author. Department of Business and
  Management Science, NHH Norwegian School of Economics. Helleveien 30,
  5045 Bergen, Norway. \texttt{hakon.otneim@nhh.no}} \\ Dag Tjøstheim\footnote{University of Bergen}}
    \preauthor{\centering\large\emph}
  \postauthor{\par}
    \date{}
    \predate{}\postdate{}
  
\newcommand{\X}{\bm{X}}
\newcommand{\Xone}{\bm{X}^{(1)}}
\newcommand{\Xtwo}{\bm{X}^{(2)}}
\newcommand{\x}{\bm{x}}

\newcommand{\tX}{\tilde{X}}

\newcommand{\Z}{\bm{Z}}
\newcommand{\z}{\bm{z}}
\newcommand{\hz}{\widehat{z}}
\newcommand{\hfz}{\widehat{\bm{z}}}
\newcommand{\hZ}{\widehat{\bm{Z}}}
\newcommand{\hnZ}{\widehat{Z}}
\newcommand{\Zone}{\bm{Z}^{(1)}}
\newcommand{\Ztwo}{\bm{Z}^{(2)}}
\newcommand{\zone}{\bm{z}^{(1)}}
\newcommand{\ztwo}{\bm{z}^{(2)}}

\newcommand{\y}{\bm{y}}
\newcommand{\Y}{\bm{Y}}
\newcommand{\hY}{\widehat{Y}}

\newcommand{\fv}{\bm{v}}
\newcommand{\fu}{\bm{u}}

\newcommand{\fC}{\bm{C}}

\newcommand{\R}{\bm{R}}
\newcommand{\hR}{\widehat{\bm{R}}}

\newcommand{\hF}{\widehat{F}}
\newcommand{\hf}{\widehat{f}}

\newcommand{\hrho}{\widehat{\rho}}
\newcommand{\frho}{\bm{\rho}}
\newcommand{\hfrho}{\widehat{\bm{\rho}}}

\newcommand{\hh}{\bm{b}}

\newcommand{\fmu}{\bm{\mu}}

\newcommand{\fSigma}{\bm{\Sigma}}
\newcommand{\hfSigma}{\widehat{\bm{\Sigma}}}
\newcommand{\halpha}{\widehat{\alpha}}
\newcommand{\fOmega}{\bm{\Omega}}
\newcommand{\fLambda}{\bm{\Lambda}}
\newcommand{\fepsilon}{\bm{\epsilon}}

\newcommand{\Jb}{\bm{J}_{\hh}}
\newcommand{\Mb}{\bm{M}_{\hh}}

\usepackage{amsthm}

\newtheorem{corollary}{Corollary}

\newcommand{\Cov}{\textrm{Cov}}
\newcommand{\E}{\textrm{E}}
 
\newcommand{\di}{\,\textrm{d}}

\usepackage{graphicx}
\usepackage{subfig}
\usepackage{bm}
\usepackage{booktabs}
\usepackage{caption}
\usepackage{nimbusmononarrow}
\usepackage[T1]{fontenc}
\usepackage{color}
\usepackage{colortbl}

\definecolor{Gray}{gray}{0.9}

\let\BeginKnitrBlock\begin \let\EndKnitrBlock\end
\begin{document}
\maketitle
\begin{abstract}
It is well known that the dependence structure for jointly Gaussian
variables can be fully captured using correlations, and that the
\textit{conditional} dependence structure in the same way can be
described using partial correlations. The partial correlation does not,
however, characterize conditional dependence in many non-Gaussian
populations. This paper introduces the local Gaussian partial
correlation (LGPC), a new measure of conditional dependence. It is a
local version of the partial correlation coefficient that characterizes
conditional dependence in a large class of populations. It has some
useful and novel properties besides: The LGPC reduces to the ordinary
partial correlation for jointly normal variables, and it distinguishes
between positive and negative conditional dependence. Furthermore, the
LGPC can be used to study departures from conditional independence in
specific parts of the distribution. We provide several examples of this,
both simulated and real, and derive estimation theory under a local
likelihood framework. Finally, we indicate how the LGPC can be used to
construct a powerful test for conditional independence, which, again,
can be used to detect Granger causality in time series.
\end{abstract}

\section{Introduction}\label{introduction}

Estimation of conditional dependence and testing for conditional
independence are extremely important topics in classical as well as
modern statistics. In the last two decades, for instance, there has been
a very intense development using conditional dependence in probabilistic
network theory. This comes in addition to conditional multivariate time
series analysis and copula analysis.

For jointly Gaussian variables, conditional dependence is measured by
the partial correlation coefficient. Given three jointly Gaussian
stochastic variables \(X_1\), \(X_2\) and \(X_3\), \(X_1\) and \(X_2\)
are conditionally independent given \(X_3\) if and only if the partial
correlation between \(X_1\) and \(X_2\) given \(X_3\) is equal to zero.
There is a rich literature on applications of the partial correlation
coefficient to Gaussian networks, path analysis and causality. The
Gaussian assumption is strict, however, and it is easy to find
non-Gaussian examples where the partial correlation function completely
fails in describing conditional dependence. We will give some explicit
examples of that in Section \ref{chap:examples}.

The purpose of this paper is to introduce a new concept for measuring
conditional dependence. This concept retains all the properties of the
ordinary partial correlation in the Gaussian case, but seeks to avoid
the weaknesses of this measure in the non-Gaussian case. We do this by
fitting a \emph{family} of Gaussian distributions to a given continuous
multivariate distribution and exploiting the simple conditioning rules
of the Gaussian distribution \emph{locally}. This approach produces a
new measure of conditional dependence, \emph{the local Gaussian partial
correlation (LGPC)} which, being directly related to the ordinary
partial correlation, is easy to interpret, and reduces to the very same
ordinary partial correlation in the Gaussian case. Moreover, it
distinguishes between positive and negative conditional dependence
whereas competing non-linear measures report only the \emph{strength} of
the conditional dependence on some non-negative scale.

The local view gives much more flexibility. It allows the conditional
dependence to be stronger or weaker in certain regions of a multivariate
distribution than in others. This is of particular interest in finance,
where description of tail behavior is important. The local aspect also
makes it possible to focus tests of conditional independence to selected
parts of the distribution, which may potentially increase the power of
the test.

The local approach has been shown to be advantageous in other areas of
statistics, such as the measurement of nonlinear dependence, density
estimation and spectral analysis, see for instance Tjøstheim and
Hufthammer (\protect\hyperlink{ref-tjostheim2013local}{2013}), Otneim
and Tjøstheim (\protect\hyperlink{ref-otneim2017locally}{2017}) and
Jordanger and Tjøstheim
(\protect\hyperlink{ref-jordanger2017nonlinear}{2017}). The present
paper, then, represents the first attempt to model conditional
dependence locally. There are several other approaches to modelling
conditional dependence and testing for conditional independence in the
non-Gaussian case in the literature. For the most part, they revolve
around test statistics in conditional independence tests. Such
statistics are usually based on measures of \emph{distance} between some
property of the sample and the corresponding property of the population
under the null hypothesis of conditional independence. As a consequence,
the conditional dependence measures are not always easy to interpret,
and can, as mentioned above, not distinguish between negative and
positive conditional dependence. We present a comprehensive set of
references to this literature in Section \ref{chap:testing}.

A quick summary of the paper is as follows: Partial correlation in the
global case can be defined in several ways, which all coincide in the
Gaussian case. We will examine this in the following section and make a
choice of a measure that is most convenient for localization. The LGPC
is defined in Section \ref{chap:lgpc} by using a distributional local
Gaussian approach, where it will be seen that the LGPC can be introduced
at several levels of computational and theoretical complexity.
Visualization is important when studying local conditional dependence,
which involves at least three scalar variables. Several simulated and a
real data example are given, in particular cases where the ordinary
global partial correlation fails completely. Estimation theory comes
next in Section \ref{chap:estimation}, whereas testing is treated in
Section \ref{chap:testing}, including a fairly comprehensive comparison
with existing tests.

\section{Conditional and partial correlation}\label{chap:conditional}

Let \(\X = (X_1, \ldots, X_p)\) be a random vector. We will in this
paper denote by \((\Xone, \Xtwo)\) a partition of \(\X\), and in our
treatment below, \(\Xone\) will consist of the first and second
component in \(\X\) so that \(\Xone = (X_1, X_2)\), and \(\Xtwo\) will
contain the remaining components: \(\Xtwo = (X_3, \ldots, X_p)\). In any
case, we will assume that the mean vector \(\fmu\) and covariance matrix
\(\fSigma\) of \(\X\) exist, and we will partition them correspondingly,
writing

\begin{equation}
\fmu = \begin{pmatrix} \fmu_1 \\ \fmu_2 \end{pmatrix}, \,\,\, \textrm{ and } \fSigma = \begin{pmatrix} \fSigma_{11} & \fSigma_{12} \\ \fSigma_{21} & \fSigma_{22}\end{pmatrix},
\label{eq:matrixpartition}
\end{equation}

where \(\fSigma_{11}\) and \(\fSigma_{22}\) are the covariance matrices
of \(\Xone\) and \(\Xtwo\) respectively. There are two main concepts of
correlation when \(\Xtwo\) is given, the \emph{partial} and the
\emph{conditional} correlation. They coincide in several joint
distributions, among them the Gaussian. We will in this section provide
some details about this distinction and explain our preference for using
the partial correlation as a starting point when defining the LGPC.

The \emph{partial} variance-covariance matrix of \(\Xone\) given
\(\Xtwo\) is

\begin{equation}
\fSigma_{11.2} = \fSigma_{11} - \fSigma_{12}\fSigma^{-1}_{22}\fSigma_{21}.
\label{eq:matrixpartial}
\end{equation}

Note that in the Gaussian case \(\fSigma_{11.2}\) is the covariance
matrix in the conditional (Gaussian) distribution of \(\Xone\) given
\(\Xtwo\). Similarly, if \(\Xone = (X_1, X_2)\), the partial correlation
between \(X_1\) and \(X_2\) is then defined as

\begin{equation}
\rho_{11.2} = \frac{\sigma_{11.2}^{(1,2)}}{\sqrt{\sigma_{11.2}^{(1,1)}\sigma_{11.2}^{(2,2)}}},
\label{eq:partialcorrelation}
\end{equation}

where \(\sigma_{11.2}^{(i,j)}\) refers to the element in position
\((i,j)\) in the partial covariance matrix \(\fSigma_{11.2}\). This, in
turn, in the Gaussian case, can be identified with the correlation
matrix in the conditional distribution of \(\Xone\) given \(\Xtwo\), and
this fact, in particular equations \eqref{eq:matrixpartial} and
\eqref{eq:partialcorrelation} will serve as the starting point for our
definition of local partial correlation.

The partial variance or covariance of \(\Xone\) given \(\Xtwo\) can also
be considered as the variance or covariance between residuals of
projections of \(X_1\) and \(X_2\) on the linear space spanned by
\(\Xtwo\),

\[\sigma_{11.2} = \Cov\left(X_i - \tX_i(\Xtwo), X_j - \tX_j(\Xtwo) \right), \qquad i,j = 1,2,\]
where
\(\tX_i(\Xtwo) = \E(X_i) + \Sigma_{X_i\Xtwo}\Sigma_{\Xtwo\Xtwo}^{-1}(\Xtwo - \E(\Xtwo))\)
is the projection of \(X_i\) on \(\Xtwo\). The expression in terms of
residuals is often taken as the starting point when defining the the
global partial correlation, see e.g. Lawrance
(\protect\hyperlink{ref-lawrance1976conditional}{1976}). Moreover, using
eq. \eqref{eq:partialcorrelation}, it can be shown, that \emph{all} of the
partial correlations between two variables \(X_i\) and \(X_j\) in a set
of variables \(V\) having correlation matrix
\(\Omega = \{\omega_{ij}\} = \{\rho_{X_iX_j}\}\) and given the other
variables in \(V\) (i.e.~given \(V\backslash\{X_i,X_j\}\)) is obtained
as

\[\rho_{X_iX_j.V\backslash X_iX_j} = -\frac{p_{ij}}{\sqrt{p_{ii}p_{jj}}},\]
where \(p_{ij}\) is element \((i,j)\) in the precision matrix
\(\fSigma^{-1}\).

The \emph{conditional} covariance of \(X_1\) and \(X_2\) given \(\Xtwo\)
is defined by

\[\Cov(X_1,X_2|\Xtwo) = \E\left((X_1 - \E(X_1|\Xtwo))(X_2 - \E(X_2|\Xtwo))|\Xtwo\right),\]
and this is the covariance between \(X_1\) and \(X_2\) in the
conditional distribution of \((X_1,X_2)\) given \(\Xtwo\). The
conditional covariance matrix can be written

\[\Sigma_{11|2} = \begin{pmatrix} \sigma_{11|\Xtwo} & \sigma_{12|\Xtwo} \\ \sigma_{21|\Xtwo} & \sigma_{22|\Xtwo}\end{pmatrix}\]
resulting in the conditional correlation

\[\rho_{12|\Xtwo} = \frac{\sigma_{12|\Xtwo}}{\sqrt{\sigma_{11|\Xtwo}\sigma_{22|\Xtwo}}},\]
which is the correlation between \(X_1\) and \(X_2\) in the conditional
distribution of \((X_1,X_2)\) given \(\Xtwo\), thus coinciding with
\(\rho_{11.2}\) of eq. \eqref{eq:partialcorrelation} in the Gaussian case.

Baba, Shibata, and Sibuya
(\protect\hyperlink{ref-baba2004partial}{2004}) give the following
result for the relationship between conditional and partial quantities
in a more general situation.

\begin{corollary}
\label{cor:baba1}
 For any random vectors $\Xone = (X_1, X_2)$ and $\Xtwo = (X_3,\ldots,X_p)$ the following two conditions are equivalent:
\begin{itemize}
    \item[(i)] $\E(\Xone|\Xtwo) = \bm{\alpha} + \bm{B}\Xtwo$ for a vector $\alpha$ and a matrix $\bm{B}$, and $\fSigma_{11|2}$ independent of $\Xtwo$.
    \item[(ii)] $\fSigma_{11.2} = \fSigma_{11|2}$ almost surely.
\end{itemize}
\end{corollary}

Either of the conditions in Corollary 1 is, according to Baba, Shibata,
and Sibuya (\protect\hyperlink{ref-baba2004partial}{2004}), valid not
only for the multivariate normal, but also for elliptical, multivariate
hypergeometric, multinomial and Dirichlet distributions. Moreover, for
the multivariate Gaussian, zero partial correlation (or conditional
correlation) is equivalent to conditional independence between \(X_1\)
and \(X_2\) given \(\Xtwo\). We will now use such results to construct a
generalized partial correlation function, which characterizes
conditional dependence in a broader class of distributions.

\section{The local Gaussian partial correlation}\label{chap:lgpc}

\subsection{Definition}\label{chap:definition}

In all of the following it is assumed that the \(X\)-variables are
continuous and have a density function. In order to localize the partial
correlation we need to look at the concept of local Gaussian
approximations. Let \(f\) be a multivariate density function. Given a
point \(\x\), we can approximate \(f\) in a neighborhood of \(\x\) by a
multivariate Gaussian density

\begin{equation} 
\psi(\x, \fv) = \frac{1}{(2\pi)^{p/2}|\fSigma(\x)|^{1/2}} \exp \left\{-\frac{1}{2}(\fv - \fmu(\x))^T\fSigma^{-1}(\x)(\fv - \fmu(\x))\right\},
\label{eq:localgaussian}
\end{equation}

where \(\x = (x_1\ldots, x_p)\), \(\fmu(\x) = \{\mu_i(\x)\}\) and
\(\fSigma(\x) = \{\sigma_{ij}(\x)\}\) for \(i,j=1,\ldots p\). Moving to
another point \(\y\), there is another (generally different) Gaussian
approximation \(\psi(\y, \fv)\). In this way we approximate \(f\) by a
family of multivariate Gaussian densities defined by a set of smooth
parameter functions \(\{\fmu(\x), \fSigma(\x)\}\), and if \(f\) is
itself a Gaussian density, then the parameter functions collapse to
constants corresponding to the true parameter values, and
\(\psi(\x) \equiv f(\x)\). Hjort and Jones
(\protect\hyperlink{ref-hjort1996locally}{1996}) provide the general
framework for estimating such parameter functions non-parametrically
from a given data set using a local likelihood procedure, and the basic
idea in the following treatment is to replace the components in the
partial covariance matrix \eqref{eq:matrixpartial} by their locally
estimated counterparts in order to obtain a local measure of conditional
dependence.

Before we get that far, however, we will introduce a useful
transformation technique that greatly simplifies estimation of the LGPC.
Otneim and Tjøstheim (\protect\hyperlink{ref-otneim2017locally}{2017};
\protect\hyperlink{ref-otneim2017conditional}{2018}) show that the
estimation of local correlations becomes easier by transforming each
\(X_i\) to a standard normal variable \(Z_i = \Phi^{-1}(U_i)\), where
\(U_i\) is a uniform variable \(U_i = F_{X_i}(X_i)\) with \(F_{X_i}\)
being the cumulative distribution function of \(X_i\). In practice, we
do not know \(F_{X_i}\), but we can instead use the empirical
distribution function
\[F_{i,n}(x) = \frac{1}{n-1}\sum_{i=1}^n 1\left(X_i \leq x\right),\]
where \(1(\cdot)\) is the indicator function, and \(n\) is the number of
observations, resulting in pseudo standard normal variables
\(Z_{i,n} = \Phi^{-1}(F_{i,n}(X_i))\). Following Otneim and Tjøstheim
(\protect\hyperlink{ref-otneim2017locally}{2017};
\protect\hyperlink{ref-otneim2017conditional}{2018}) we then choose to
simplify the locally Gaussian approximation \eqref{eq:localgaussian} on
the marginally normal scale by considering the marginals separately, and
by fixing \(\mu_i(\z) = 0\) and \(\sigma_i(\z) = 1\) for
\(i=1,\ldots, p\), which leaves a matrix
\(\R(\z) = \{\rho_{ij}(\z)\}_{i<j}\) of local correlations to be
estimated. An alternative, used by Jordanger and Tjøstheim
(\protect\hyperlink{ref-jordanger2017nonlinear}{2017}) is to allow
\(\mu_i(\z)\) and \(\sigma_i(\z)\) to depend on \(\z\).

In the following, we will not always distinguish between \(Z_i\) and
\(Z_{i,n}\). In fact, we conclude our asymptotic analysis in Section
\ref{chap:asymptotic} by showing that under certain technical
conditions, the error made by estimating \(\R(\z)\) using the
empirically transformed variables \(Z_{i,n}\), instead of \(Z_i\), is
smaller in the limit than the estimation error made when estimating the
local correlations themselves.

Define the random vector \(\Z\) by this transformation of
\(\X = (\Xone, \Xtwo) = (X_1, X_2, \ldots, X_p)\) to marginal standard
normality:

\begin{equation}
\Z = \Big(\Phi^{-1}\left(F_{X_1}(X_1)\right), \Phi^{-1}\left(F_{X_2}(X_2)\right), \ldots, \Phi^{-1}\left(F_{X_p}(X_p)\right)\Big).
\label{eq:trans}
\end{equation}

Assume then that the probability density function of \(\Z\) can be
written on the local simplified Gaussian form (cf.~Otneim and Tjøstheim
(\protect\hyperlink{ref-otneim2017locally}{2017};
\protect\hyperlink{ref-otneim2017conditional}{2018}))

\begin{equation}
f_{\Z}(\z) = \psi(\z, R(\z)) = \frac{1}{|2\pi\R(\z)|^{1/2}}\exp\left\{-\frac{1}{2}\z^T\R^{-1}(\z)\z\right\},
\label{eq:lgdeapprox}
\end{equation}

where \(\R(\z)\) is the \emph{local correlation matrix} with
\(R(\z) = \left\{\rho_{ij}(z_i, z_j)\right\}_{i\leq j}\) and
\(\rho_{ii}(z_i) = 1\), and where the means and standard deviations have
been set and fixed to \(0\) and \(1\) correspondingly, as indicated
above. This means that we can use the terms \emph{local correlation} and
\emph{local covariance} interchangeably within this family of
distributions. We will in this paper also refer to \(\X\) and its
probability density function \(f_{\X}\) as being on the \(x\)-scale, and
to \(\Z\) and its probability density function \(f_{\Z}\) as being on
the \(z\)-scale. For further discussion of the simplified
\(z\)-representation we refer to Otneim and Tjøstheim
(\protect\hyperlink{ref-otneim2017locally}{2017}).

Denote by \((\Zone, \Ztwo)\) the partitioning of \(\Z\) corresponding to
the partitioning \((\Xone, \Xtwo)\) of \(\X\). A natural definition of
the \emph{local} partial covariance matrix of \(\Zone|\Ztwo\) is the
local version of eq. \eqref{eq:matrixpartial}:

\begin{equation}
\fSigma_{11.2}(\z) = \R_{11}(\zone) - \R_{12}(\z)\R^{-1}_{22}(\ztwo)\R_{21}(\z),
\label{eq:matrixlocalpartial}
\end{equation}

which is a \(2\times2\) matrix, and we define the local Gaussian partial
correlation \(\alpha(\z)\) of \(\Zone\) given \(\Ztwo\) in accordance
with the ordinary (global) partial correlation provided by eq.
\eqref{eq:partialcorrelation}:

\begin{equation}
\alpha(\z) = \frac{\Big\{\fSigma_{11.2}(\z)\Big\}_{12}}{\Big\{\fSigma_{11.2}(\z)\Big\}_{11}\Big\{\fSigma_{11.2}(\z)\Big\}_{22}},
\label{eq:definition}
\end{equation}

or, if \(\Ztwo = Z_3\) is scalar,

\begin{equation}
\alpha(\z) = \rho_{12|3}(z_1, z_2|z_3) = \frac{\rho_{12}(z_1, z_2) - \rho_{13}(z_1,z_3)\rho_{23}(z_2,z_3)}{\sqrt{1 - \rho^2_{13}(z_1,z_3)}\sqrt{1 - \rho^2_{23}(z_2, z_3)}}.
\label{eq:scalardefinition}
\end{equation}

The monotone relation between the \(x\)-scale and the \(z\)-scale is
given in eq. \eqref{eq:trans}. The \(z\)-representation is in a sense
analogous to a copula representation which is based directly on uniform
variables \(U_i = F^{-1}_{X_i}(X_i)\) (sometimes referred to as the
\(u\)-scale), but avoids the problems of unbounded densities on bounded
support that often occur in common copula models. It is of course
possible to introduce an LGPC \(\alpha(x)\) directly on the \(x\)-scale,
but this representation is in many ways harder to handle both
computationally and asymptotically. For a multivariate Gaussian
distribution, we have that \(\alpha_X(\x) = \alpha_Z(\z) = \alpha\). In
the remainder of this paper, we will mainly write in terms of the
\(z\)-representation using the LGPC \(\alpha(\z) = \alpha_Z(\z)\), but
when we write the local partial correlation between \(X_1\) and \(X_2\)
given \(\X_3 = \x_3\) at the point \((x_1,x_2,x_3)\), this is simply
\(\alpha(\z)\) inserted with \(z_i = \Phi^{-1}(F_{X_i}(x_i))\),
\(i=1,\ldots,p\).

\subsection{Properties}\label{properties}

The local Gaussian partial correlation is clearly closely related to the
partial correlation between jointly normally distributed variables. We
see this also from the following properties that we will establish next:

\begin{enumerate}
\def\labelenumi{\arabic{enumi}.}
\tightlist
\item
  The LGPC \(\alpha(\z)\) satisfies \(-1 \leq \alpha(\z) \leq 1\).
\item
  If \(\X\) is jointly normally distributed, then the LGPC coincides
  with the ordinary (global) partial, and thus conditional correlation.
\item
  For stochastic vectors having joint density function on \(z\)-scale of
  the form \eqref{eq:lgdeapprox}, the LGPC \(\alpha(\z)\) is identically
  equal to zero if and only if \(X_1\) and \(X_2\) are conditionally
  independent given \(\Xtwo\). Note that \(\alpha(\z) \equiv 0\) if and
  only if \(\alpha(\x) \equiv 0\). \label{prop:char}
\item
  The LGPC is invariant with regard to a set of monotone transformations
  \(\Y = \bm{h}(\X) = (h_1(X_1), \ldots, h_p(X_p))\).
\end{enumerate}

Property 1 is trivially true if \(\R(\z)\) is a valid correlation
matrix. By removing the \(\z\)-dependence in the local correlations, it
follows immediately from eq. \eqref{eq:lgdeapprox} and from the results
referred to in Section \ref{chap:conditional} that property 2 holds.

To see that conditional independence between \(X_1\) and \(X_2\) given
\(\Xtwo\) is equivalent to \(\alpha(\z) \equiv 0\), or equivalently
\(\alpha(\x) \equiv 0\), we need to follow a few simple steps that would
also work in the global Gaussian case. Working on the standard normal
\(z\)-scale and assuming that \((z_1, z_2, \z_3)\) is jointly locally
normally distributed having density function \(f_{\Z}(\z)\) as given by
\eqref{eq:lgdeapprox}, it follows from Otneim and Tjøstheim
(\protect\hyperlink{ref-otneim2017conditional}{2018}) that we can
calculate conditional distributions in the same way as in the global
Gaussian case. This is indeed an important advantage of using the
Gaussian family as local approximant as compared to other families of
parametric distributions. For example, the conditional density of
\(z_1|\z_3\) at the point \(z_1\) is given by

\[f_{z_1|\z_3}(z_1) = \frac{1}{|2\pi\sigma_{1|3}(z_1)|^{1/2}}\times\exp \left\{-\frac{1}{2}\left(\frac{z_1 - \mu_{1|3}(z_1)}{\sigma_{1|3}(z_1)}\right)^2\right\},\]
where

\begin{equation}
\mu_{1|3}(z_1) = \R_{12}(z_1, \z_3)\R_{22}(\z_3)^{-1}\z_3 \,\,\, \textrm{ and } \,\,\, \sigma_{1|3}(z_1) = \fSigma_{Z_1|\Z_3},
\label{eq:localconditionalpar}
\end{equation}

where the latter expression is defined in the same way as
\eqref{eq:matrixlocalpartial}. The conditional density
\(f_{Z_2|\Z_3}(z_2|\z_3)\) is defined in the same way. If \(Z_1\) and
\(Z_2\) are conditionally independent given \(\Z_3\), then

\begin{align*}
f_{Z_1,Z_2|\Z_3}(z_1, z_2|\z_3) &= f_{Z_1|\Z_3}(z_1|\z_3)f_{Z_2|\Z_3}(z_2|\z_3) \\
&= \frac{1}{2\pi\sqrt{1-\sigma_{1|3}(z_1)}\sqrt{1-\sigma_{2|3}(z_2)}} \\ 
&\qquad \qquad \times\exp\left\{-\frac{1}{2}\left(\frac{z_1 - \mu_{1|3}(z_1)}{\sigma_{1|3}(z_1)}\right)^2\right\}\exp\left\{-\frac{1}{2}\left(\frac{z_2 - \mu_{2|3}(z_2)}{\sigma_{2|3}(z_2)}\right)^2\right\},
\end{align*}

and we at once identify the conditional density of \((Z_1,Z_2)|\Z_3\) as
another Gaussian distribution, but without any cross-term involving
\(z_1z_2\), which then implies that the off-diagonal element in the
partial covariance matrix \eqref{eq:matrixlocalpartial} is identically
equal to zero. We see immediately from our definition
\eqref{eq:definition} that the LGPC is also identically equal to zero.

The converse statement also follows by looking at the conditional
density of \((Z_1, Z_2|\Z_3)\). Let \(f_{\Z}(\z)\) be on the locally
Gaussian form \eqref{eq:lgdeapprox}. Again, we have from Otneim and
Tjøstheim (\protect\hyperlink{ref-otneim2017conditional}{2018}) that the
conditional density of \((Z_1, Z_2|\Z_3)\) is locally associated with a
Gaussian distribution, with local conditional mean vector and local
conditional covariance matrix given by expressions corresponding to
\eqref{eq:localconditionalpar}. The off-diagonal in the local conditional
covariance matrix is zero if \(\alpha(\z) \equiv 0\), or equivalently
\(\alpha(\x) \equiv 0\), in which case one may factorize the joint
conditional density of \((Z_1, Z_2|\Z_3)\) into two factors, one
depending only on \(z_1\), and one depending only on \(z_2\). Hence,
\(Z_1\) and \(Z_2\) are conditionally independent given \(\Z_3\).

Let the transformation \(\Y = \bm{h}(\X)\) define the \(y\)-scale in the
same way as the transformation \eqref{eq:trans} defines the \(z\)-scale.
We define the LGPC in terms of the marginally standard normal
\(\Z\)-variables, which for the \(\Y\)-variables can be calculated as

\[\Z = \Big(\Phi^{-1}\left(F_{Y_1}(Y_1)\right), \ldots, \Phi^{-1}\left(F_{Y_p}(Y_p)\right)\Big),\]
but if \(h_j(\cdot)\) is monotone for all \(j = 1,\ldots,p\),

\[F_{Y_j}(y_j) = P(h_j(X_j) \leq y_j) = P(X_j \leq h_j^{-1}(y_j)) = P(X_j \leq x_j) = F_{X_j}(x_j),\]
where \(x_j\) is the point on the \(x\)-scale corresponding to the point
\(y_j\) on the \(y\)-scale. The \(\Z\)-variables are the same for the
stochastic variables \(\X\) and \(\Y = \bm{h}(\X)\). Their LGPC-function
\(\alpha(\z)\) must therefore be the same as well according to our
definition in the preceding section.

\section{Estimating conditional dependence via the
LGPC}\label{chap:estimation}

\subsection{Estimation by local likelihood}\label{ll}

We see from the defining equations \eqref{eq:matrixlocalpartial} and
\eqref{eq:definition} that the basic building blocks for the \emph{local
Gaussian partial correlation} are the \emph{local Gaussian correlation
functions}, that populate the local correlation matrix \(\R(\z)\) in the
density function \(f_{\Z}\) in \eqref{eq:lgdeapprox}. Consider the random
vector \(\X = (X_1, X_2, \ldots, X_p)\) having joint probability density
function (pdf) \(f(\x)\), and its transformed counterpart
\(\Z = \left(\Phi^{-1}(F_1(X_1)), \ldots, \Phi^{-1}(F_p(X_p))\right)\),
on the marginally standard normal \(z\)-scale. The relation between the
pdf of \(\X\) and the pdf of \(\Z\) is given by

\begin{equation}
f_{\X}(\x) = f_{\Z}\left(\Phi^{-1}\left(F_1(x_1)\right), \ldots, \Phi^{-1}\left(F_p(x_p)\right)\right)\prod_{i=1}^p 
\frac{f_i(x_i)}{\phi\left(\Phi^{-1}\left(F_i(x_i)\right)\right)},
\label{eq:fxfz}
\end{equation}

where \(f_i(\cdot)\), \(i = 1,\ldots, p\) are the marginal density
functions of \(\X\), and \(\phi(\cdot)\) is the standard normal density
function. Otneim and Tjøstheim
(\protect\hyperlink{ref-otneim2017locally}{2017}) provide more details
of this construction in the context of multivariate density estimation.
Let \(\X_1, \ldots, \X_n\) be a random sample identically distributed as
\(\X\) and construct the pseudo standard normal observations
\(\hZ_1, \ldots, \hZ_n\) as

\begin{equation}
\hZ_i = \left(\Phi^{-1}\left(\hF_1(X_{1i})\right), \ldots, \Phi^{-1}\left(\hF_p(X_{pi})\right)\right),
\label{eq:pseudoobservations}
\end{equation}

where \(\X_i = (X_{1i}, X_{2i}, \ldots, X_{p_i})\), and \(\hF_j(\cdot)\)
is an estimate of the marginal distribution function of \(X_j\), for
example the empirical distribution function. Next, we must produce an
estimate \(\hR(\z)\) of the local correlation matrix, and we will
consider two variations of this task in the following sub-sections. It
is then natural to estimate the local Gaussian partial covariance matrix
as

\begin{equation}
\hfSigma_{11.2}(\z) = \hR_{11}(\z) -  \hR_{12}(\z)\hR_{22}^{-1}(\z)\hR_{21}(\z),
\label{eq:localpartialcov}
\end{equation}

and if we assume that \(X_1\) and \(X_2\) constitute the two first
elements in \(\X\), we estimate the local Gaussian partial correlation
between \(X_1\) and \(X_2\) given \(\Xtwo\) on the \(z\)-scale as

\begin{equation}
\widehat\alpha(\z) = \frac{\Big\{\hfSigma_{11.2}(\z)\Big\}_{12}}{\Big\{\hfSigma_{11.2}(\z)\Big\}_{11}\Big\{\hfSigma_{11.2}(\z)\Big\}_{22}}.
\label{eq:estimate}
\end{equation}

A corresponding value of \(\halpha(\x)\) at the point
\(\x = \widehat F^{-1}(\Phi(z))\) is obtained by inserting
\(\z = \Phi^{-1}(\widehat F(\x))\) in \eqref{eq:estimate}. The \emph{curse
of dimensionality} dictates that nonparametric estimates of multivariate
regression and density functions quickly deteriorates as the number of
variables increases. This is also true for local correlation surfaces,
but it can be alleviated somewhat by introducing heavier smoothing
towards the global Gaussian model.

Otneim and Tjøstheim (\protect\hyperlink{ref-otneim2017locally}{2017})
propose a different strategy. By estimating each local correlation
\(\rho_{ij}(\z)\), \(1\leq i<j\leq p\), sequentially in the bivariate
problem involving only the variables \(Z_i\) and \(Z_j\), they
effectively impose a pairwise dependence structure in order to
circumvent the curse of dimensionality. This is indeed possible when
calculating the LGPC as well, and it can be done for an arbitrary
dimension \(p\). It is described in more detail in Section
\ref{chap:bivariate}.

In the case that \(p = 3\), and all variables involved are scalars, we
do provide in this paper the empirical option and corresponding
estimation theory to fit a full trivariate normal distribution locally
to \((\Zone, \Ztwo) = (Z_1,Z_2,Z_3)\) without any structural
simplifications. This constitutes a novelty in the local Gaussian
literature, and we present the procedure in the following sub-section.

\subsubsection{\texorpdfstring{Estimation of \(\R(\z)\) when \(p = 3\)
and \(\Xtwo\) is a
scalar}{Estimation of \textbackslash{}R(\textbackslash{}z) when p = 3 and \textbackslash{}Xtwo is a scalar}}\label{chap:trivariate-full}

If dim\((\X) = 3\), then dim\((\Xtwo)=1\), and \(\R(\z)\) is a
\(3\times3\) symmetric matrix of local correlation functions having
arguments \(\z = (z_1,z_2,z_3)\):
\[\R(\z) = \begin{pmatrix} 1 & \rho_{12}(\z) & \rho_{13}(\z) \\ \rho_{12}(\z) & 1 & \rho_{23}(\z) \\ \rho_{13}(\z) & \rho_{23}(\z) & 1\end{pmatrix},\]
containing three parameter functions that must be estimated from data.
Hjort and Jones (\protect\hyperlink{ref-hjort1996locally}{1996}) provide
the local likelihood framework that we need to perform this task. Based
on the sample \eqref{eq:pseudoobservations}, we estimate \(\R(\z)\) by
fitting the parametric family \(\psi(z, \R)\), defined in
\eqref{eq:lgdeapprox}, locally to the density \(f_{\Z}(\z)\) by maximizing
the \emph{local likelihood function}

\begin{equation}
\hR(\z) = \arg\max_{\R} n^{-1} \sum_{i=1}^nK_{\hh}(\hZ_i - \hfz)\log\psi(\hZ_i, \R) - \int K_{\hh}(\y - \hfz)\psi(\y, \R)\,\textrm{d}\y
\label{eq:locallik3}
\end{equation}

in each point \(\z\), where
\(K_{\hh}(\x) = |\hh|^{-1/2}K(\hh^{-1/2}\x)\), \(K\) is a non-negative
and radially symmetric kernel function that satisfies
\(\int K(\x)\,\textrm{d}\x=1\), and \(\hh\) is a diagonal \(3\times3\)
matrix of bandwidths that serve as smoothing parameters. The estimate
\(\hf_{\Z}(\z) = \psi(\z, \hR(\z))\) aims at minimizing its locally
weighted Kullback-Leibler distance to \(f_{\Z}(\z)\), and we refer to
Hjort and Jones (\protect\hyperlink{ref-hjort1996locally}{1996}) and
Tjøstheim and Hufthammer
(\protect\hyperlink{ref-tjostheim2013local}{2013}) for much more details
about this construction, and to Section \ref{chap:asymptotic} for a
review of relevant estimation theory.

Finally, we obtain the estimated LGPC \(\halpha(\z)\) by plugging the
estimated local correlations \(\hR(\z)\) into equations
\eqref{eq:localpartialcov} and \eqref{eq:estimate} and a corresponding value
of \(\halpha(\x)\) at the point \(\x = \widehat F^{-1}(\Phi(\z))\) by
inserting \(\z = \Phi^{-1}({\widehat F(\x)})\).

\subsubsection{\texorpdfstring{Estimation of \(\R(\z)\) when \(\Xtwo\)
is a
vector}{Estimation of \textbackslash{}R(\textbackslash{}z) when \textbackslash{}Xtwo is a vector}}\label{chap:bivariate}

As mentioned above, the complexity of the estimation problem
\eqref{eq:locallik3} increases sharply with the number of variables
involved, much in the same way as other nonparametric estimation methods
suffer under the curse of dimensionality. Otneim and Tjøstheim
(\protect\hyperlink{ref-otneim2017locally}{2017}) suggest a way to
circumvent this issue when estimating local correlation matrices. Their
basic idea is that local correlations in \(\R(\z)\) are modeled as
functions of their corresponding \emph{pair} of variables only:

\begin{equation}
\R(\z) = \left\{\rho_{jk}(z_j, z_k)\right\}_{j<k}, \qquad j,k = 1, 2, \ldots, p,
\label{eq:loccormat}
\end{equation}

which reduces the estimation of the \(p\)-variate correlation functions
\(\rho_{jk}(\z)\) to a series of bivariate estimation problems. We
suggest to use this approach also when modelling partial dependence when
the number of variables in \(X\) is bigger than three. This allows
nonlinear dependence between variables to be approximated only by the
pairwise structure \eqref{eq:loccormat}, while still being computationally
tractable. An analogue to the pairwise approximation is the additive
approximation in nonparametric regression. See Otneim and Tjøstheim
(\protect\hyperlink{ref-otneim2017locally}{2017}) for a brief
discussion.

We turn again to Hjort and Jones
(\protect\hyperlink{ref-hjort1996locally}{1996}) for the means to
estimate the simplified version of \(\R(\z)\). We transform the
observations to the \(z\)-scale as in \eqref{eq:pseudoobservations}, but
now estimate the individual components in \(\R(\z)\) sequentially,
taking only into consideration the \emph{pair} of variables in question:

\begin{equation}
\hrho_{jk}(\hz_j, \hz_k) = \arg\max_{\rho_{jk}} n^{-1} \sum_{i=1}^nK_{\hh}(\hZ_i - \hfz)\log(\hZ_i, \rho_{jk}) - \int K_{\hh}(\y - \hfz)\psi(\y, \rho_{jk})\,\textrm{d}\y, 
\label{eq:locallik2}
\end{equation}

where all running variables and samples are bivariate subsets
corresponding to the indices \((j,k)\), \(\psi(\cdot, \rho)\) is the
bivariate version of \eqref{eq:lgdeapprox}, and \(\hh\) now is a
\(2\times2\) diagonal matrix of bandwidths. After estimating all local
correlations in this way, we proceed to calculate the LGPC using
equations \eqref{eq:localpartialcov} and \eqref{eq:estimate} as above.

\subsection{Asymptotic theory}\label{chap:asymptotic}

Equations \eqref{eq:matrixlocalpartial}, \eqref{eq:definition} and their
empirical counterparts \eqref{eq:localpartialcov} and \eqref{eq:estimate}
demonstrate clearly that the LGPC is nothing more than a deterministic
function of the local correlation matrix, in the same way as the
ordinary partial correlation is a function of the ordinary correlation
matrix. The asymptotic behavior of \(\halpha(\z)\) can thus be derived
directly from the asymptotic behavior of \(\hR(\z)\), which, for the
most part, has been established in earlier works, see e.g. Tjøstheim and
Hufthammer (\protect\hyperlink{ref-tjostheim2013local}{2013}) and Otneim
and Tjøstheim (\protect\hyperlink{ref-otneim2017locally}{2017}). We will
therefore in the following mostly refer to those results, and rather
direct our focus to some details that are different in our context. We
will start by assuming that the marginal distribution functions
\(F_1, \ldots, F_p\) are known, and then state two results on the joint
asymptotic normality of the estimated local correlation matrix
\(\hR(\z)\) under a mixing condition. We will then apply the delta
method in order to derive the limiting distribution of the estimated
LGPC, and finally show that we may replace \(F_1,\ldots,F_p\) with their
empirical counterparts \(\hF_1,\ldots,\hF_p\) without changing these
asymptotic results.

Consider first the full trivariate fit described in Section
\ref{chap:trivariate-full} with \(\z = (z_1,z_2,z_3)\). We estimate the
local correlation matrix \(\R(\z)\) by maximizing the local likelihood
function in \eqref{eq:locallik3}, which we denote by \(L_n(\R(\z), \z)\),
with the only exception that we, for now, assume the true transformation
between the \(x\)- and the \(z\)-scale to be known. For a fixed matrix
of bandwidths \(\hh\), denote by \(\R_{\hh}(z)\) the local correlation
that satisfies, as \(n \rightarrow \infty\),

\begin{equation}
\nabla L_n(\R_{\hh}, \z) \rightarrow \int K_{\hh}(\y - \z)\fu(\y, \R_{\hh}(\z))\{f_{\Z}(\y) - \psi(\y, \R_{\hh})\}\,\textrm{d}\y = 0,
\label{eq:popfixedh3}
\end{equation}

where \(f_{\Z}(\z) = f_{Z_1,Z_2,Z_3}(z_1,z_2,z_3)\) is the density
function of \(\Z\), and \(\fu(\cdot, \rho_{\hh})\) is the column vector
of local score functions \(\nabla \log\psi(\cdot, \rho_{\hh})\), where
the gradient is taken with respect to the parameters \(\rho_{12}(\z)\),
\(\rho_{13}(\z)\) and \(\rho_{23}(\z)\). Denote by
\(\frho = \frho(\z) = \{\rho_{12}(\z), \rho_{13}(\z), \rho_{23}(\z)\}\)
the vector of local correlations. The joint limiting distribution of
\(\hfrho\) (for convenience only stated on the \(z\)-scale below) is
given by the following result as \(n \rightarrow \infty\) and
\(\hh \rightarrow \bm{0}\):

\vspace{.5cm}

\BeginKnitrBlock{theorem}
\protect\hypertarget{thm:loccor3}{}{\label{thm:loccor3} }Let \(\{\Z\}_n\) be
observations on the marginally standard normally distributed random
vector \(\Z\) having joint density \(f_{\Z}(\z)\). Assume that the
following conditions hold:

\begin{enumerate}
\def\labelenumi{\arabic{enumi}.}
\tightlist
\item
  For any sequence of bandwidth matrices \(\hh_n\) tending to zero
  element-wise, there exists for the trivariate marginally standard
  Gaussian vector \(\Z\) a unique set of local correlations
  \(\frho_{\hh}(\z)\) that satisfies \eqref{eq:popfixedh3}, and there
  exists a \(\frho_0(\z)\) such that
  \(\frho_{\hh}(\z) \rightarrow \frho_0(\z)\).
\item
  \(\{\Z\}_n\) is \(\alpha\)-mixing with mixing coefficients satisfying
  \(\sum_{m\geq1}m^{\lambda}\alpha(m)^{1-2/\delta} < \infty\) for some
  \(\lambda > 1-2/\delta\) and \(\delta>2\).
\item
  \(n\rightarrow\infty\), and each bandwidth \(b\) tends to zero such
  that
  \(nb^{\frac{\lambda+2-2/\delta}{\lambda + 2/\delta}} = O(n^{\epsilon_0})\)
  for some constant \(\epsilon_0 > 0\).
\item
  In a given point \(\z\), the parameter space \(\Theta\) for each local
  correlation \(\rho(\z)\) is a compact subset of \((-1,1)\).
\item
  The kernel function \(K(\cdot)\) sasisfies
  \(\sup_{\z}|K(\z)| < \infty\),
  \(\int |K(\y)|\,\textrm{d}\y < \infty\),
  \(\partial/\partial z_iK(\z) <\infty\), and
  \(\lim_{z_i \rightarrow \infty}|z_iK(\z)| = 0\) for \(i = 1,2,3\).
\end{enumerate}

Then, by writing \(b_{n1} = b_{n2} = b_{n3} = b \rightarrow 0\) assuming
they all converge to zero at the same rate,

\begin{equation}
\sqrt{nb^5}\Jb\Mb^{-1/2}\left(\hfrho_n - \frho_0\right) \stackrel{\mathcal{L}}{\rightarrow} \mathcal{N}(\bm{0}, \bm{I}),
\label{eq:an3}
\end{equation}

where \(\bm{I}\) is the \(3\times3\) identity matrix,

\begin{align*}
\Jb &= \int K_{\hh}(\y - \z)\fu(\y, \frho_{\hh}(\z))\fu^T(\y, \frho_{\hh}(\z))\psi(\y,\frho_{\hh}(\z))\,\textrm{d}\y \\
& \qquad\qquad\qquad\qquad - \int K_{\hh}(\y - \z)\nabla\fu(\y,\frho_{\hh}(\z))\Big\{f_{\Z}(y) - \psi(\y,\frho_{\hh}(\z))\Big\}\textrm{d}\y,
\end{align*}

and

\begin{align*}
\Mb &= b^3\int K^2_{\hh}(\y - \z)\fu(\y, \frho_{\hh}(\z))\fu^T(\y,\frho_{\hh}(\z))f_{\Z}(\y)\,\textrm{d}\y \\
& \qquad\qquad - b^3\int K_{\hh}(\y - \z)\fu(\y, \frho_{\hh}(\z))f_{\Z}(\y)\,\textrm{d}\y \int K_{\hh}(\y - \z)\fu^T(\y, \frho_{\hh}(\z))f_{\Z}(\y)\,\textrm{d}\y.
\end{align*}
\EndKnitrBlock{theorem}

Theorem \ref{thm:loccor3} and its assumptions above are slight
variations of the assumptions and the results presented by Hjort and
Jones (\protect\hyperlink{ref-hjort1996locally}{1996}), and in
particular Theorem 3 by Tjøstheim and Hufthammer
(\protect\hyperlink{ref-tjostheim2013local}{2013}). The latter reference
contains a detailed argument concerning the convergence rate of local
likelihood estimates in multi-parameter problems, such as ours. Their
analysis of the asymptotic behavior of \(\Jb\) and \(\Mb\) reveals that
the local parameter estimates converge more slowly than, for example,
the trivariate kernel density estimator (that converges as
\(1/(nb^3)\)), and it is a straightforward exercise to modify their
proof to suit our particular situation, which reveals that the three
local correlations converge jointly in distribution at the rate
\(1/(nb^5)\). Furthermore, Tjøstheim and Hufthammer
(\protect\hyperlink{ref-tjostheim2013local}{2013}) demonstrate that an
analytic expression of the leading term in the covariance matrix
\(\Jb^{-1}\Mb(\Jb^{-1})^T\) is not practically available. In
applications we may circumvent this problem by using the bootstrap, see
also Lacal and Tjøstheim
(\protect\hyperlink{ref-lacal2018estimating}{2018}), or by approximating
\(\Jb\) and \(\Mb\) using empirical moments corresponding to the
integrals defining them. We refer to Appendix \ref{app-loccor3} for some
details regarding the proof of Theorem \ref{thm:loccor3}.

Moving on to the higher dimensional case, we will see below that the
pairwise estimation structure that we describe in Section
\ref{chap:bivariate} simplifies the asymptotic analysis of \(\hR(\z)\)
considerably. This result is also available in the literature already,
so we modify the notation slightly, and reproduce Theorem 3 in Otneim
and Tjøstheim (\protect\hyperlink{ref-otneim2017conditional}{2018}). For
a pair of variables \((X_j,X_k)\) and its corresponding pair of
transformed variables \((Z_j, Z_k)\) (the relation between the \(x\)-
and \(z\)-scale is again assumed to be known at this stage). We then
estimate the single local correlation
\(\rho_{jk}(x_j,x_k) = \rho_{ij}(z_j,z_k)\) by maximizing the local
likelihood function in \eqref{eq:locallik2}, denoted by
\(L_n(\rho_{ij}, \z)\), For a fixed matrix of bandwidths \(\hh\), now a
diagonal \(2\times2\)-matrix, define \(\rho_{\hh}\) in a similar way as
above as the solution to

\begin{equation}
\frac{\partial L_n(\rho_{\hh}, \z)}{\partial\rho} \rightarrow \int K_{b}(\y - \z)u(\y, \rho_{\hh})\{f_{jk}(\y) - \psi(\y, \rho_{\hh})\}\,\textrm{d}\y = 0
\label{eq:popfixedh}
\end{equation}

as \(n \rightarrow \infty\), but now \(f_{jk}(\z)\) is the joint density
of \((Z_j, Z_k)\), \(u(\cdot, \rho_{\hh})\) is the local score function
\(\partial \log\psi(\cdot, \rho_{\hh})/\partial\rho\), and \(K(\cdot)\)
and \(\psi(\cdot)\) are bivariate versions of the kernel function and
the simplified Gaussian density associated with \eqref{eq:lgdeapprox}
respectively. Denote by \(\frho = \{\rho_{jk}\}_{j<k}\) the vector of
all local correlations defined between all pairs of variables in \(\X\).
The joint limiting distribution of \(\hfrho(\z)\) is then given by the
following result (Otneim and Tjøstheim
\protect\hyperlink{ref-otneim2017locally}{2017}):

\vspace{.5cm}

\BeginKnitrBlock{theorem}
\protect\hypertarget{thm:loccor}{}{\label{thm:loccor} }Let \(\{\Z\}_n\) be
observations on the marginally standard normally distributed random
vector \(\Z\), and for each pair of variables \((Z_j, Z_k)\),
\(j<k\leq p\), assume that the following conditions hold:

\begin{enumerate}
\def\labelenumi{\arabic{enumi}.}
\tightlist
\item
  For any sequence of bandwidth matrices \(\hh = \hh_n\) tending to zero
  element-wise, there exists for the bivariate marginally standard
  Gaussian vector \((Z_j, Z_k)\) a unique \(\rho_{\hh}(\z)\) that
  satisfies \eqref{eq:popfixedh}, and there exists a \(\rho_0(\z)\) such
  that \(\rho_{\hh}(\z) \rightarrow \rho_0(\z)\).
\item
  For each pair \((j,k)\) of variables, with \(1\leq j\leq p\),
  \(1\leq k\leq p\), \(j\neq k\), \(\{(Z_j,Z_k)\}_n\) is
  \(\alpha\)-mixing with mixing coefficients satisfying
  \(\sum_{m\geq1}m^{\lambda}\alpha(m)^{1-2/\delta} < \infty\) for some
  \(\lambda > 1-2/\delta\) and \(\delta>2\).
\item
  \(n\rightarrow\infty\), and each bandwidth \(b\) tends to zero such
  that
  \(nb^{\frac{\lambda+2-2/\delta}{\lambda + 2/\delta}} = O(n^{\epsilon_0})\)
  for some constant \(\epsilon_0 > 0\).
\item
  For a given location \(\z\), the parameter space \(\Theta\) for each
  local correlation \(\rho(\z)\) is a compact subset of \((-1,1)\).
\item
  The kernel function \(K(\cdot)\) sasisfies
  \(\sup_{\z}|K(\z)| < \infty\),
  \(\int |K(\y)|\,\textrm{d}\y < \infty\),
  \(\partial/\partial z_iK(\z) <\infty\), and
  \(\lim_{z_i \rightarrow \infty}|z_iK(\z)| = 0\) for \(i = 1,2\).
\end{enumerate}

Then, by letting \(b_n^2\) (assuming identical convergence rates) mean
the product \(b_{nj}b_{nk}\) corresponding to the pairs of variables
defining the components in \(\frho\),

\begin{equation*}
\sqrt{nb_n^2}\left(\hfrho(\z) - \frho_0(\z)\right) \stackrel{\mathcal{L}}{\rightarrow} \mathcal{N}(\bm{0}, \fOmega),
\end{equation*}

where \(\fOmega\) is a diagonal matrix with components

\begin{equation}
\fOmega^{(\ell,\ell)} = 
\frac{f_{\ell}(\z_{\ell})\int K^2(\y_{\ell})\,\textrm{d}\y_{\ell}}
{u^2(\z_{\ell},\rho_{0,\ell}(\z_{\ell}))}
\psi^2(\z_{\ell}, \rho_{0,\ell}(\z_{\ell})),
\label{eq:asymcov}
\end{equation}

where \(\ell = 1,\ldots,p(p-1/2)\) runs over all pairs of variables, and
\(f_{\ell}(\cdot)\) is the bivariate marginal density function of the
pair \(\Z_{\ell}\).
\EndKnitrBlock{theorem}

\vspace{.5cm}

For a proof, we refer to Otneim and Tjøstheim
(\protect\hyperlink{ref-otneim2017conditional}{2018}). There are two
important differences between Theorems \ref{thm:loccor3} and
\ref{thm:loccor}. First of all, we notice in Theorem \ref{thm:loccor}
that the convergence rate is equal to the usual nonparametric rate of
\(1/nb^2\): this is because we estimate the local correlations
sequentially based on pairs of variables. The second difference is that
the leading term of the asymptotic covariance matrix in Theorem
\ref{thm:loccor} is easy to write up analytically, as we have done in
equation \eqref{eq:asymcov}.

We can now calculate the limiting distribution of the LGPC by means of
the delta method. Denote by
\(g: \mathbb{R}^{p(p-1)/2} \rightarrow \mathbb{R}\) the translation
\eqref{eq:definition} from the vector of local correlations between the
components in \(\Z\) (or, equivalently, \(\X\)) to the local partial
correlation \(\alpha(\z)\) between \(Z_1\) and \(Z_2\) given
\(Z_3, \ldots, Z_p\). From the expression of our estimate
\eqref{eq:estimate} we have that \(\halpha(\z) = g(\hfrho(\z))\), so we
use the delta method to see that

\begin{equation}
\sqrt{nb_n^m}\left(\halpha(\z) - \alpha(\z)\right) \stackrel{\mathcal{L}}{\rightarrow} \mathcal{N}\left(0, \nabla g(\frho)^T\fLambda\nabla g(\frho)\right),
\label{eq:limitlgpc}
\end{equation}

where \(m\) is equal to 2 or 5, and \(\fLambda\) is either equal to the
leading term of \(\Jb^{-1}\Mb(\Jb^{-1})^T\), or \(\fOmega\), depending
on whether we use the full trivariate locally Gaussian fit described in
Section \ref{chap:trivariate-full} or the sequentially pairwise
simplification described in Section \ref{chap:bivariate}. In any case,
it is only a matter of basic differentiation to work out an expression
for \(\nabla g(\cdot)\), and this task has been deferred to the
Appendix, section \ref{app-asvar}.

Finally, we present the limiting distribution of the LGPC between
\(Z_1\) and \(Z_2\) given \(Z_3, \ldots, Z_p\) when the components in
\(\Z = (Z_1, \ldots, Z_p)\) are replaced by marginally standard normal
\emph{pseudo-observations} as defined in equation
\eqref{eq:pseudoobservations}. The following result ensures that Theorems
\ref{thm:loccor3} and \ref{thm:loccor}, and thus eq. \eqref{eq:limitlgpc},
still holds.

\vspace{.5cm}

\BeginKnitrBlock{theorem}
\protect\hypertarget{thm:pseudo}{}{\label{thm:pseudo} }Under the conditions
in Theorems \ref{thm:loccor3} and \ref{thm:loccor}, assume further that
the kernel function \(K(\cdot)\) has bounded support. Replacing the
marginally standard normal vector \(\Z\) with the approximately
marginally standard normal vector \(\hZ\) as defined in eq
\eqref{eq:limitlgpc}, does not change the conclusions in Theorems
\ref{thm:loccor3} and \ref{thm:loccor}.
\EndKnitrBlock{theorem}

\BeginKnitrBlock{proof}
{}See the Appendix, Section \ref{proofpseudo}.
\EndKnitrBlock{proof}

The practicality of the other conditions of Theorems \ref{thm:loccor3}
and \ref{thm:loccor} have been discussed in more detail by Otneim and
Tjøstheim (\protect\hyperlink{ref-otneim2017conditional}{2018}).

\subsection{Examples}\label{chap:examples}

\subsubsection{Simulated examples}\label{chap:simulated-examples}

\begin{figure}
\subfloat[$n = $ 100\label{fig:gaussian-example-plots1}]{\includegraphics[width=0.31\linewidth]{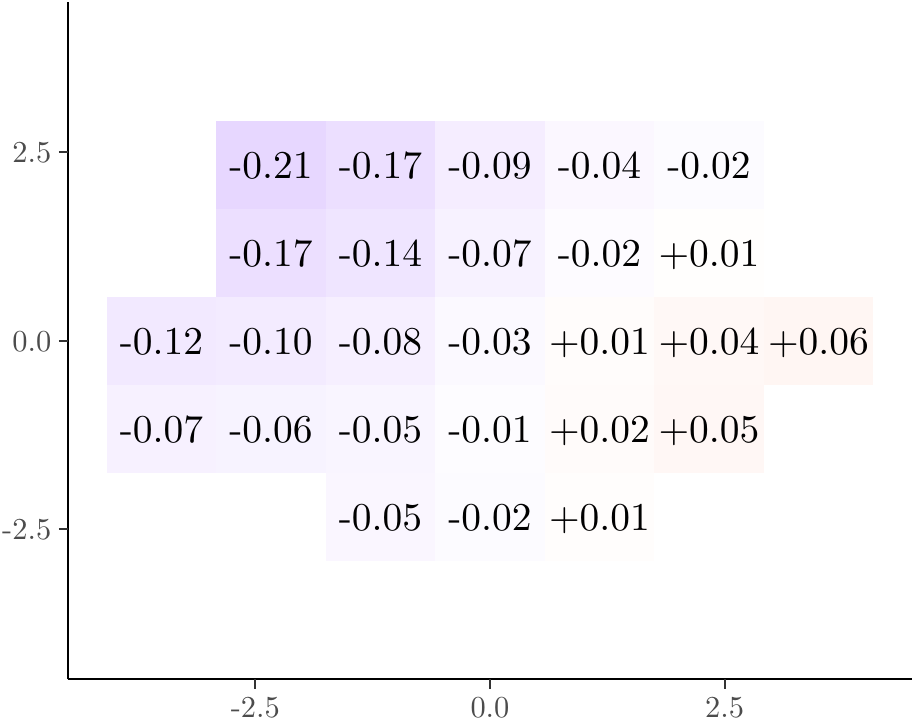} }\subfloat[$n = $ 500\label{fig:gaussian-example-plots2}]{\includegraphics[width=0.31\linewidth]{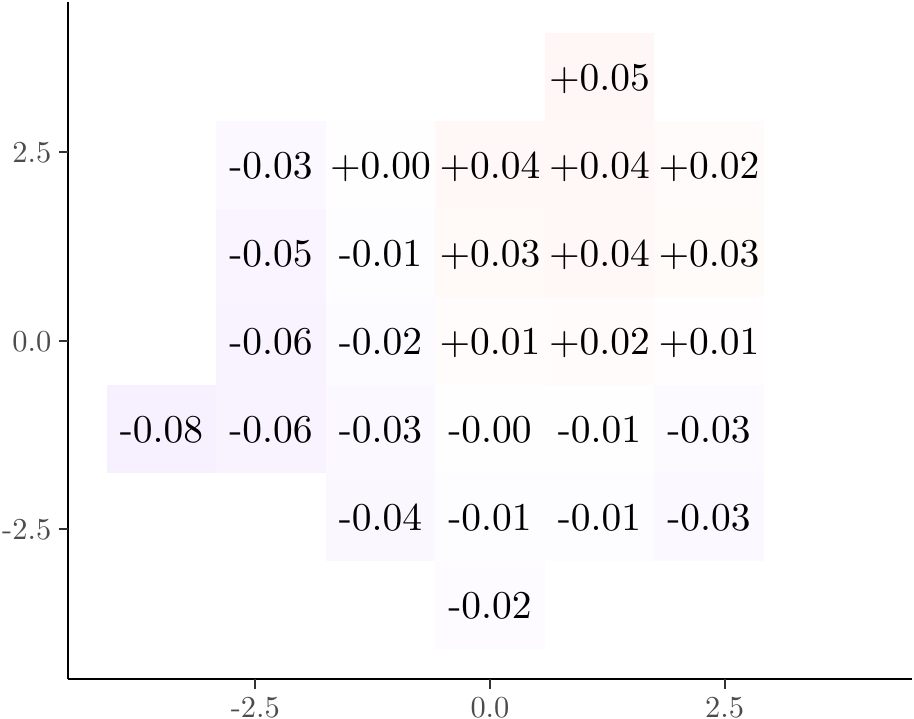} }\subfloat[$n = $ 2000\label{fig:gaussian-example-plots3}]{\includegraphics[width=0.31\linewidth]{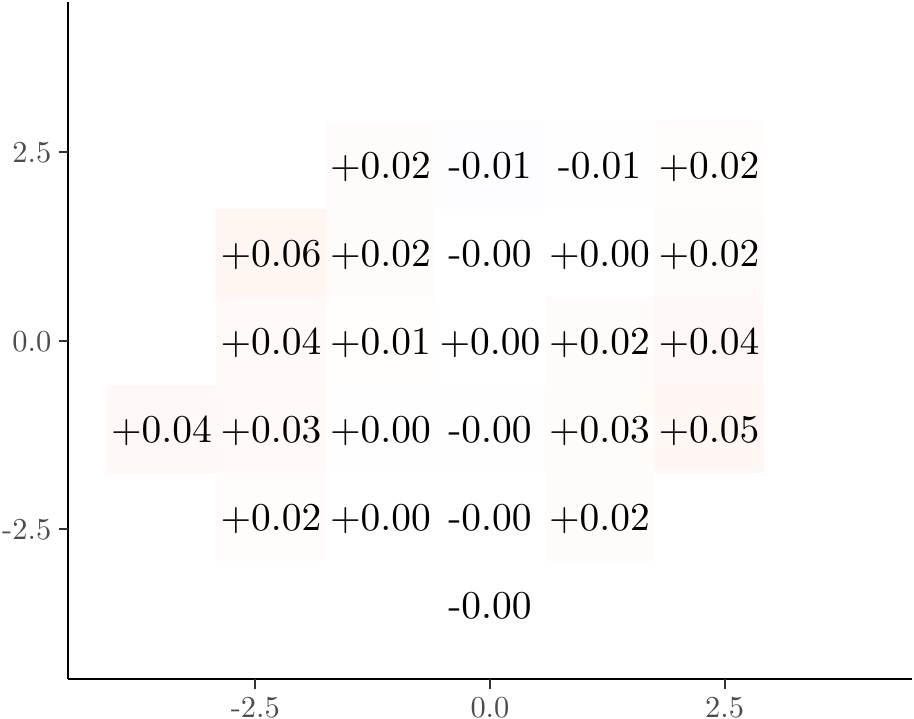} }\caption{Estimated local Gaussian partial correlation between $X_1$ and $X_2$ given $X_3 = 0$, where $(X_1, X_2, X_3)$ is a jointly standard normally distributed vector. The fully trivariate model of Section \ref{chap:trivariate-full} is used.}\label{fig:gaussian-example-plots}
\end{figure}

We will in this section provide some demonstrations on how the estimated
LGPC can be used to reveal nonlinear departures from conditional
independence. In the first and second example we use the pairwise
simplification of Section \ref{chap:bivariate}. In the third example, we
employ the fully trivariate model of Section \ref{chap:trivariate-full},
and with exception of the third simulated example, very similar results
are obtained for both estimation methods. All plots are presented on the
\(x\)-scale.

First, we provide some reference pictures from the simplest situation
imaginable. Let \(X_1\), \(X_2\) and \(X_3\) be independent standard
(and thus also jointly) normal variables. Since \(X_1\) is conditionally
independent from \(X_2\) given \(X_3\) (written \(X_1 \perp X_2 | X_3\))
in this particular case, we also know from property 3 in Section
\ref{properties} that the LGPC between \(X_1\) and \(X_2\) is zero
everywhere.

In Figure \ref{fig:gaussian-example-plots} we see the estimated LGPC
between \(X_1\) and \(X_2\) given \(X_3=0\) for three samples, having
sample size 100, 500 and 2000 respectively, mapped out on a grid, where
blue colors indicate negative local partial correlation, and red colors
indicate positive local partial correlation. These dependence maps are
inspired by Berentsen, Kleppe, and Tjøstheim
(\protect\hyperlink{ref-berentsen2014introducing}{2014}), and all
computations in this paper may be reproduced by following instructions
in the online appendix. In this and all other examples in this paper we
select bandwidths based on a simple plug-in formula that follows
naturally from classical asymptotic arguments, \(b = cn^{-1/9}\) for the
full trivariate fit, and \(b = cn^{-1/6}\) for the bivariate
simplification, where the constant \(c\) controls the amount of
smoothing and must be chosen appropriately based on the task at hand.
Håkon Otneim (\protect\hyperlink{ref-otneim2016multivariate}{2016})
argues that \(c=1.75\) is a good choice within the realm of density
estimation, and we will in the next section see that this choice also
gives good power when testing for conditional independence in many
instances, but also that even smaller values may be beneficial in
others. For the visual display of conditional dependence maps, however,
we tend to prefer more smoothing, and in Figure
\ref{fig:gaussian-example-plots} we have used \(c=4\). We will typically
calculate the LGPC at different levels of smoothing during the initial
exploration of data. Otneim and Tjøstheim
(\protect\hyperlink{ref-otneim2017locally}{2017}) also suggest a cross
validation algorithm for bandwidth selection in the density estimation
context.

As expected, \(\widehat\alpha(\x)\) is close to zero in all three plots
in Figure \ref{fig:gaussian-example-plots}.

\begin{figure}
\subfloat[Estimated local Gaussian partial correlation between $X_1$ \newline and $X_2$ given $X_3=0$ in the structural model $X_2 = X_1^2 + X_3.$\label{fig:structural-plot1}]{\includegraphics[width=0.49\linewidth]{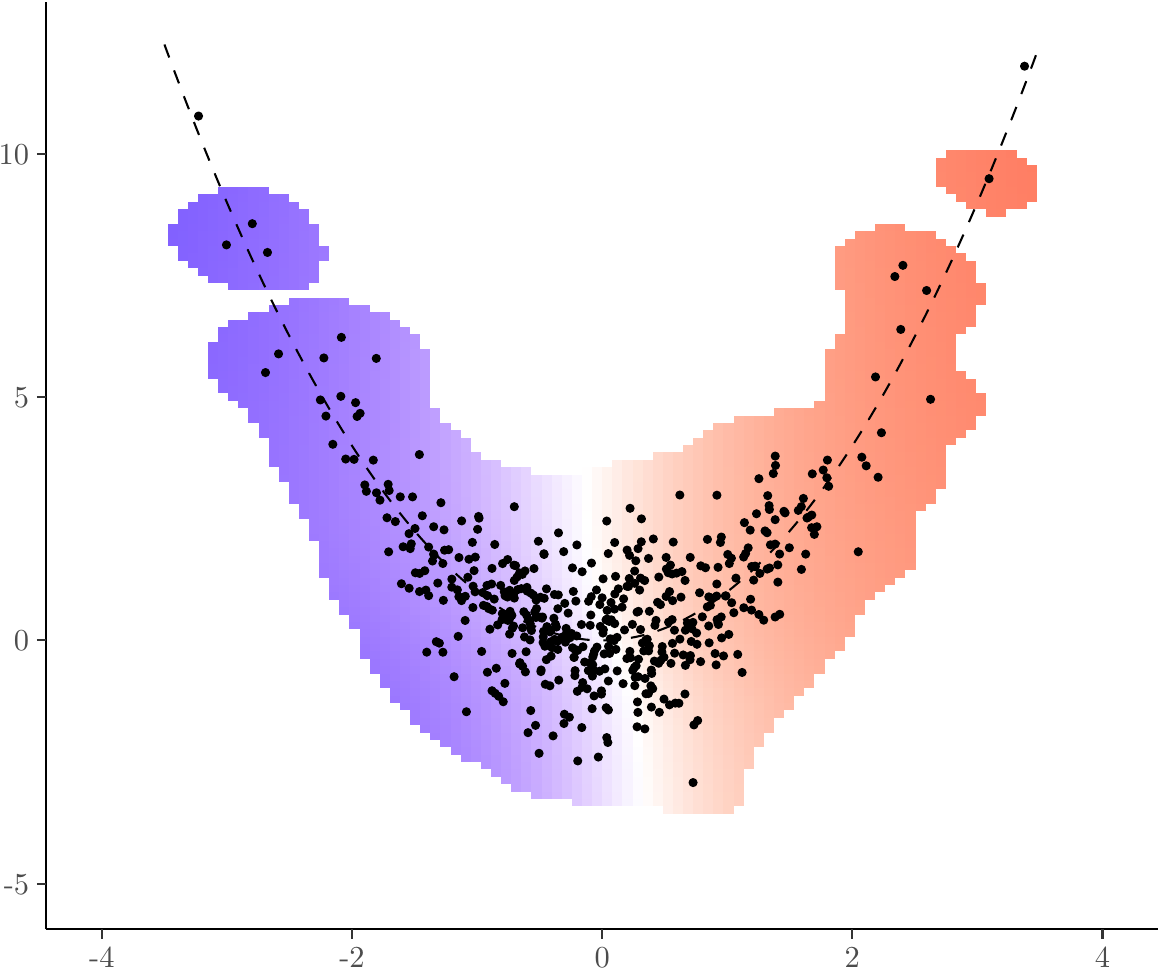} }\subfloat[The estimated LGPC along the curve $X_2 = X_1^2$ with 95\% confidence band.\label{fig:structural-plot2}]{\includegraphics[width=0.49\linewidth]{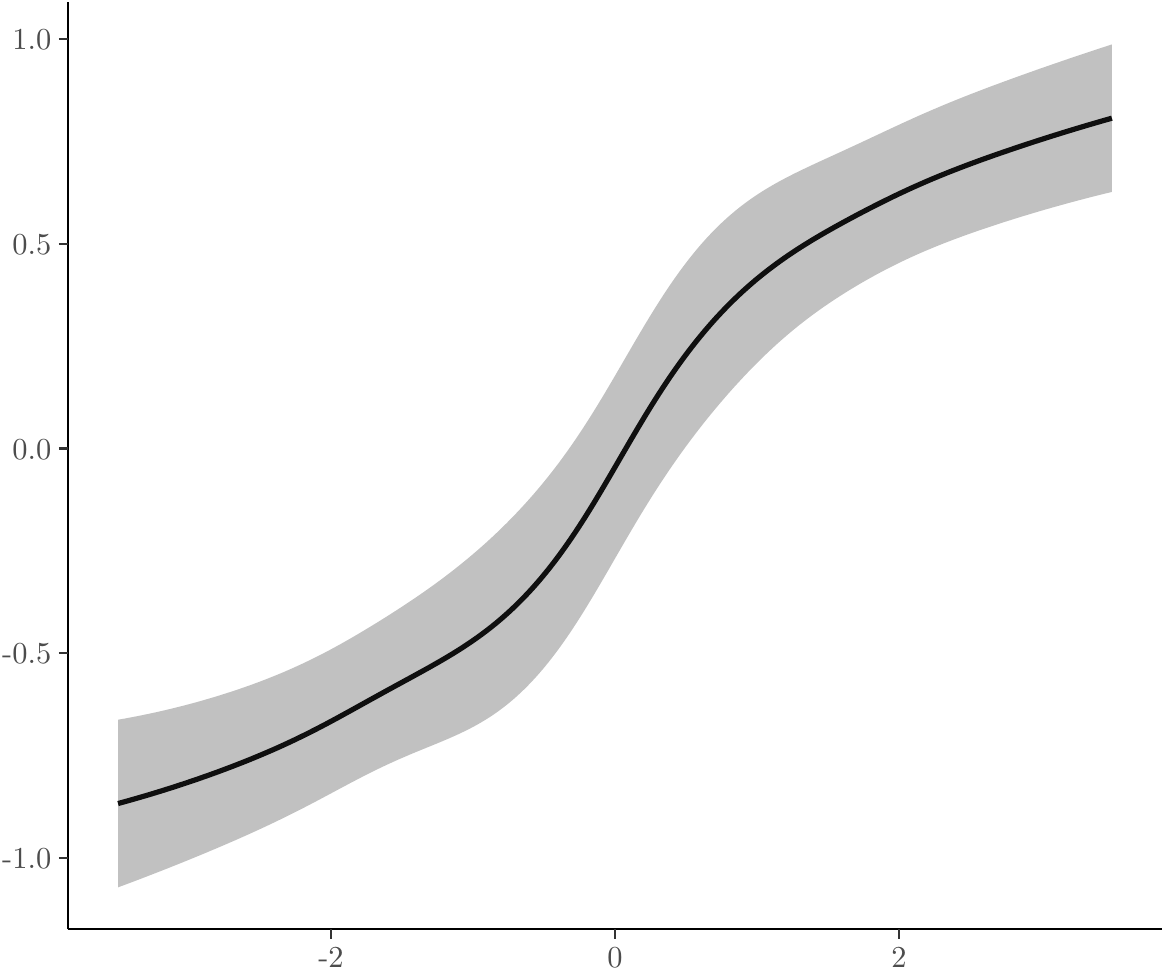} }\caption{The structural model. The fully trivariate model of Section \ref{chap:trivariate-full} is used.}\label{fig:structural-plot}
\end{figure}

We have noted before that the ordinary partial correlation coefficient
characterizes conditional dependence between jointly normal variables.
It is not hard, on the other hand, to construct examples in which strong
nonlinear relationships remain completely undetected by the partial
correlation. Consider, for example, the following structural equation;

\begin{equation}
X_2 = X_1^2 + X_3,
\label{eq:structural}
\end{equation}

where we observe all components in \((X_1, X_2, X_3)\). There is,
obviously, a strong dependence between \(X_1\) and \(X_2\), and
furthermore, this dependence is deterministic when conditioning on
\(X_3 = x_3\). It is well known, however, that \(X_2\) and
\(X_1^2 + x_3\) are uncorrelated if \(\E(X_1) = \E(X_1^3) = 0\). The
LGPC easily reveals conditional dependence between \(X_1\) and \(X_2\)
in this case. Let us generate \(n = 500\) independent observations each
from \(X_1 \sim \mathcal{N}(0,1)\) and \(X_3 \sim \mathcal{N}(0,1)\) and
calculate \(X_2\) by \eqref{eq:structural}. The sample partial correlation
between \(X_1\) and \(X_2\) is -0.037 in this case, but the LGPC, on the
other hand, which in this case is calculated using the simplified
pairwise structure defined in Section \ref{chap:bivariate}, is displayed
in Figure \ref{fig:structural-plot1} along with the observations. The
LGPC indicates strong departures from conditional independence on the
plane defined by \(X_3 = 0\). Indeed, we identify a region of
\emph{negative conditional relationship} on the left hand side of the
plot, which makes good sense because small values of \(X_2\) is
typically observed together with large values of \(X_1\) and vice versa.
The opposite phenomenon is clearly visible in the right hand half of the
plot. In Figure \ref{fig:structural-plot2} we have plotted the estimated
LGPC along the curve \(X_2 = X_1^2\) (that is indicated as a dashed line
i Figure \ref{fig:structural-plot1}), along with a 95\% confidence band,
calculated using the limiting distribution defined in Theorem
\ref{thm:loccor}. The full trivariate fit presented in Section
\ref{chap:trivariate-full} gives a very similar picture.

\begin{figure}
\subfloat[Observed pairs $(X_1, X_2)$ from the conditionally \newline Gaussian model.\label{fig:simex2-plots1}]{\includegraphics[width=0.49\linewidth]{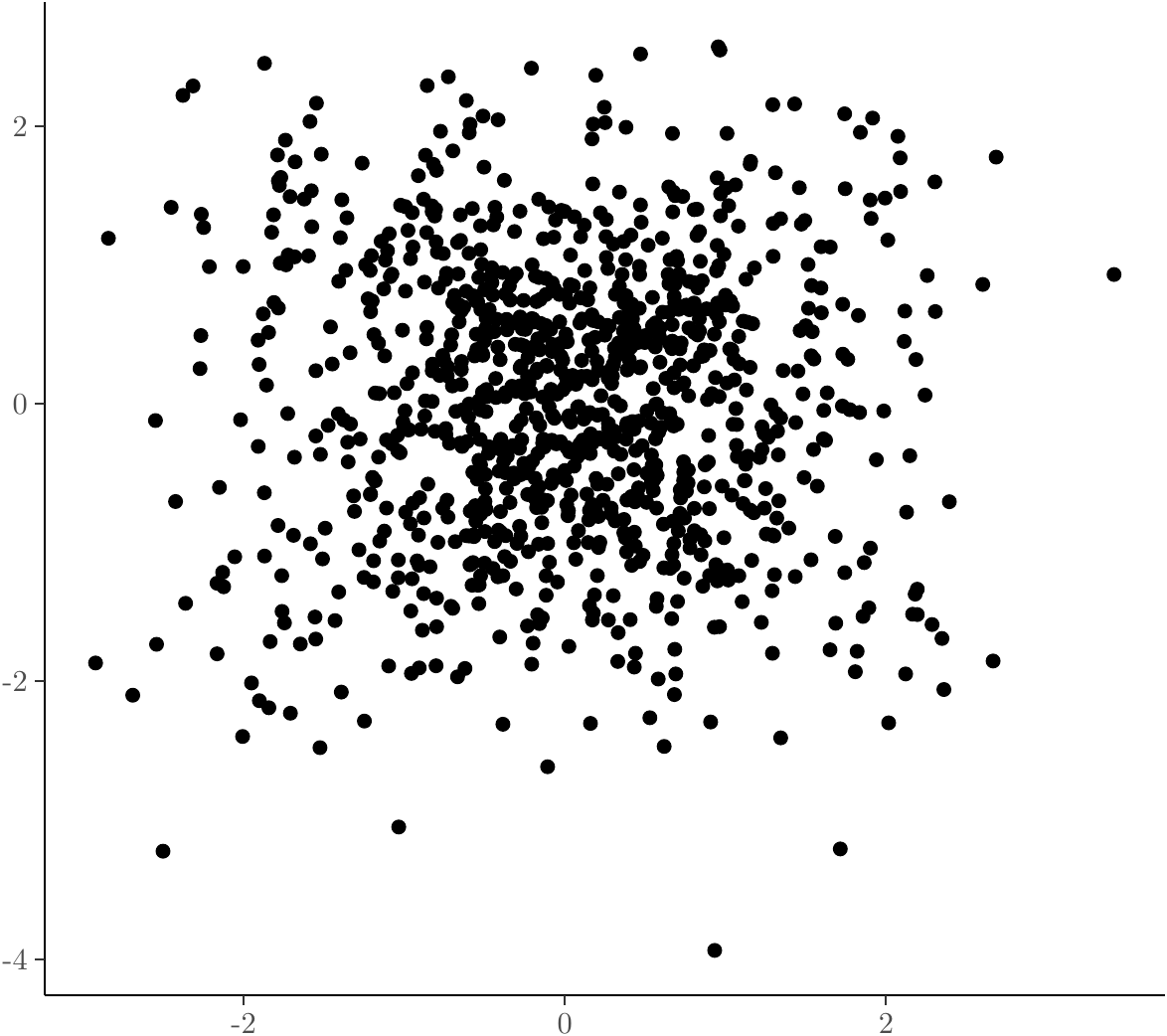} }\subfloat[The estimated LGPC between $X_1$ and $X_2$ given $\rho = -0.9$.\label{fig:simex2-plots2}]{\includegraphics[width=0.49\linewidth]{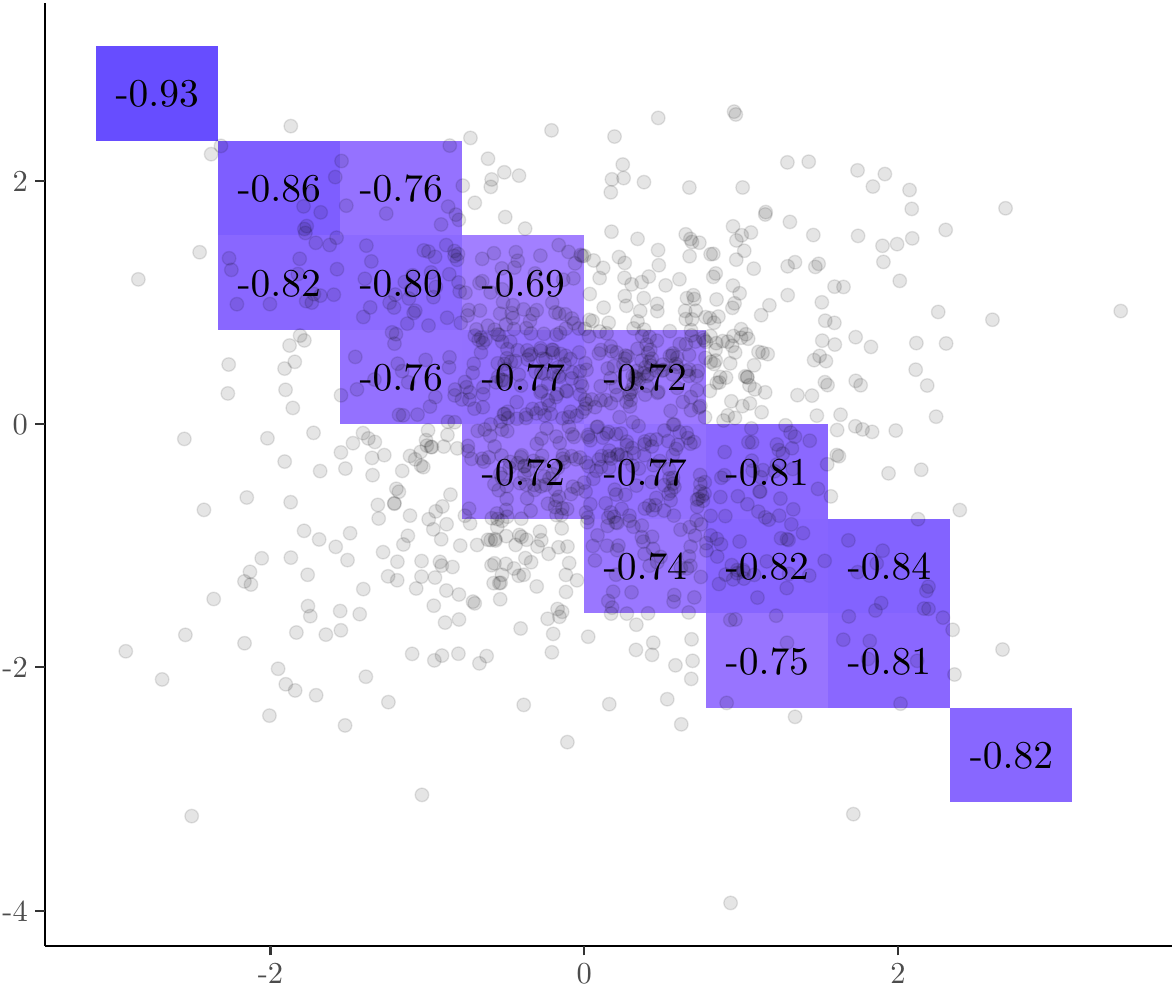} }\newline\subfloat[The estimated LGPC between $X_1$ and $X_2$ given $\rho = 0$\label{fig:simex2-plots3}]{\includegraphics[width=0.49\linewidth]{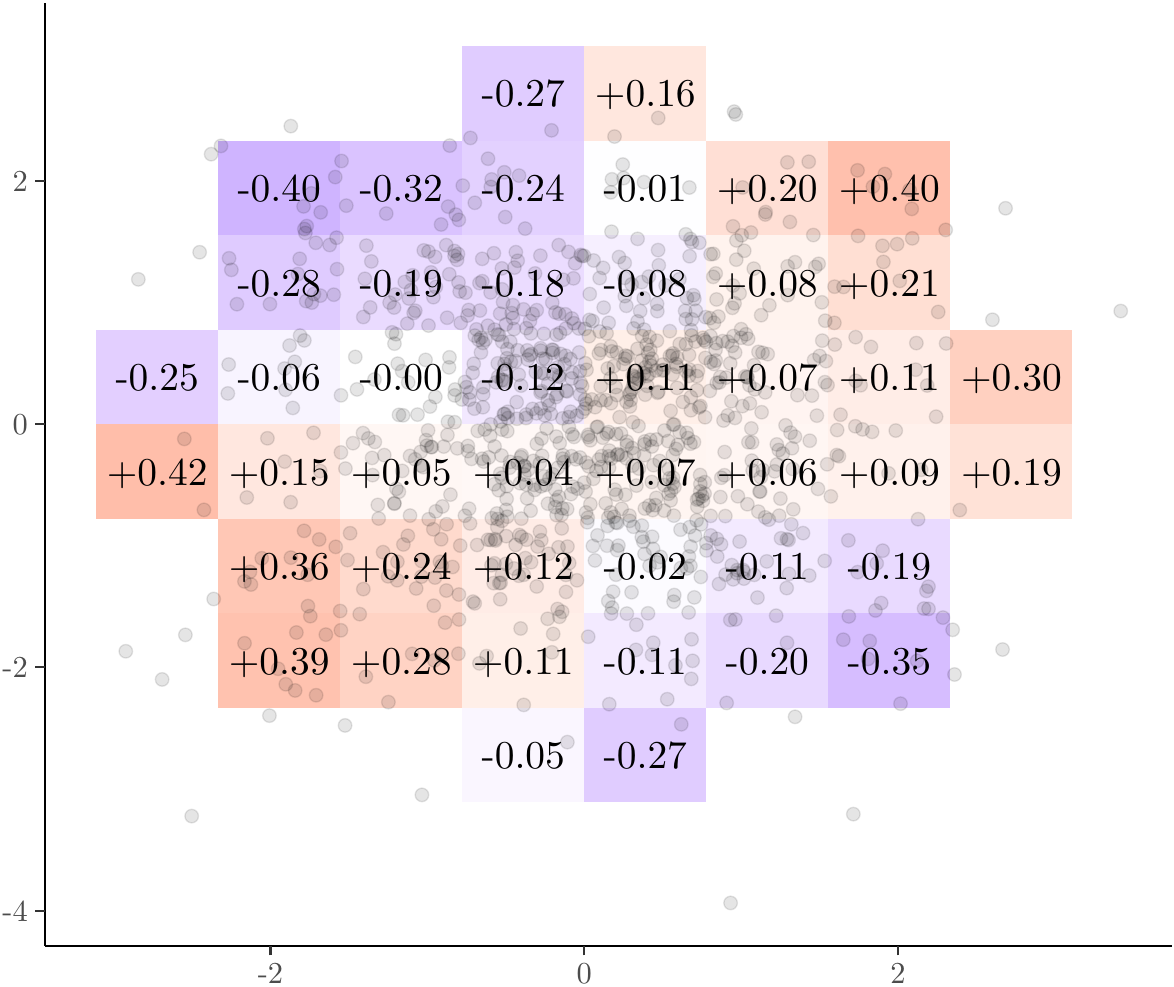} }\subfloat[The estimated LGPC between $X_1$ and $X_2$ given $\rho = 0.9$\label{fig:simex2-plots4}]{\includegraphics[width=0.49\linewidth]{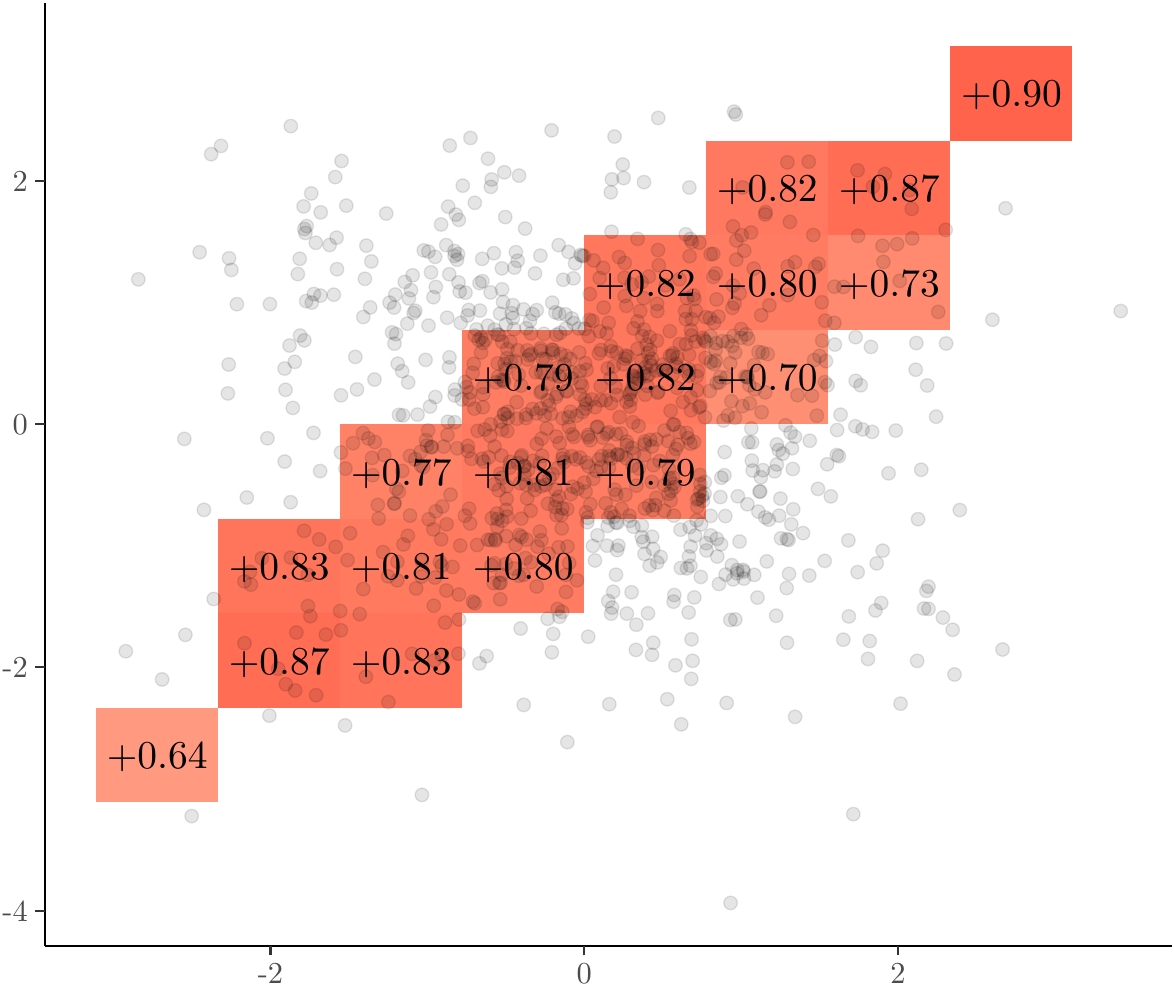} }\caption{The conditionally Gaussian model}\label{fig:simex2-plots}
\end{figure}

This points to an important feature of the LGPC: It is able to
\emph{distinguish between positive and negative conditional
relationships}, which, to our knowledge, has until now not been possible
beyond the linear and jointly Gaussian setting using the ordinary
partial correlation. Our approach also allows exploration of
conditionally different dependence patterns across different levels of
the conditioning variable. Let us demonstrate how this works in another
constructed example: Generate \(X_3 = \rho \sim U(-1,1)\), and then
generate \((X_1, X_2)\) from the bivariate Gaussian distribution having
standard normal marginals, and correlation coefficient equal to
\(\rho\). We observe \(\X = (X_1, X_2, \rho)\) and seek to visualize the
dependence between \(X_1\) and \(X_2\) conditional on \(\rho\) using the
LGPC.

We see the results from this exercise in Figure \ref{fig:simex2-plots}.
In panel (a), we generate \(n=1000\) simulated pairs \((X_1, X_2)\) from
this model. In panel (b)-(d) we see the estimated LGPC plotted over
suitable grids, and where the conditioning variable \(X_3 = \rho\) has
been fixed at the respective values \(-0.9\), \(0\) and \(0.9\), and we
see clearly how the dependence between \(X_1\) and \(X_2\) changes in
these cases. In this particular example it is straightforward to see
that \(X_1\) and \(X_2\) are uncorrelated and that the ordinary partial
correlation between \(X_1\) and \(X_2\) given \(X_3\) is equal to zero.
Furthermore, we note that the pairwise simplification defined in Section
\ref{chap:bivariate} would also not be able to measure the
\emph{conditional} relationship between \(X_1\) and \(X_2\) given
\(X_3\) in this case, as we see clearly from eq.
\eqref{eq:scalardefinition}, because \(X_1\) and \(X_2\) are both
marginally independent from \(X_3\). This means that the two pairwise
correlations \(\rho_{13}(z_1,z_3)\) and \(\rho_{23}(z_2,z_3)\) are equal
to zero. This example also shows that the form of the LGPC can depend
very strongly on the value of the conditioning variable.

\subsubsection{Empirical example: Granger
causality}\label{empirical-example-granger-causality}

\begin{figure}
\subfloat[Local partial dependence map of $V_t^*$ vs. $R_{t-1}$ on the \newline plane defined by $V_{t-1}^* = 0$\label{fig:realplots1}]{\includegraphics[width=0.49\linewidth]{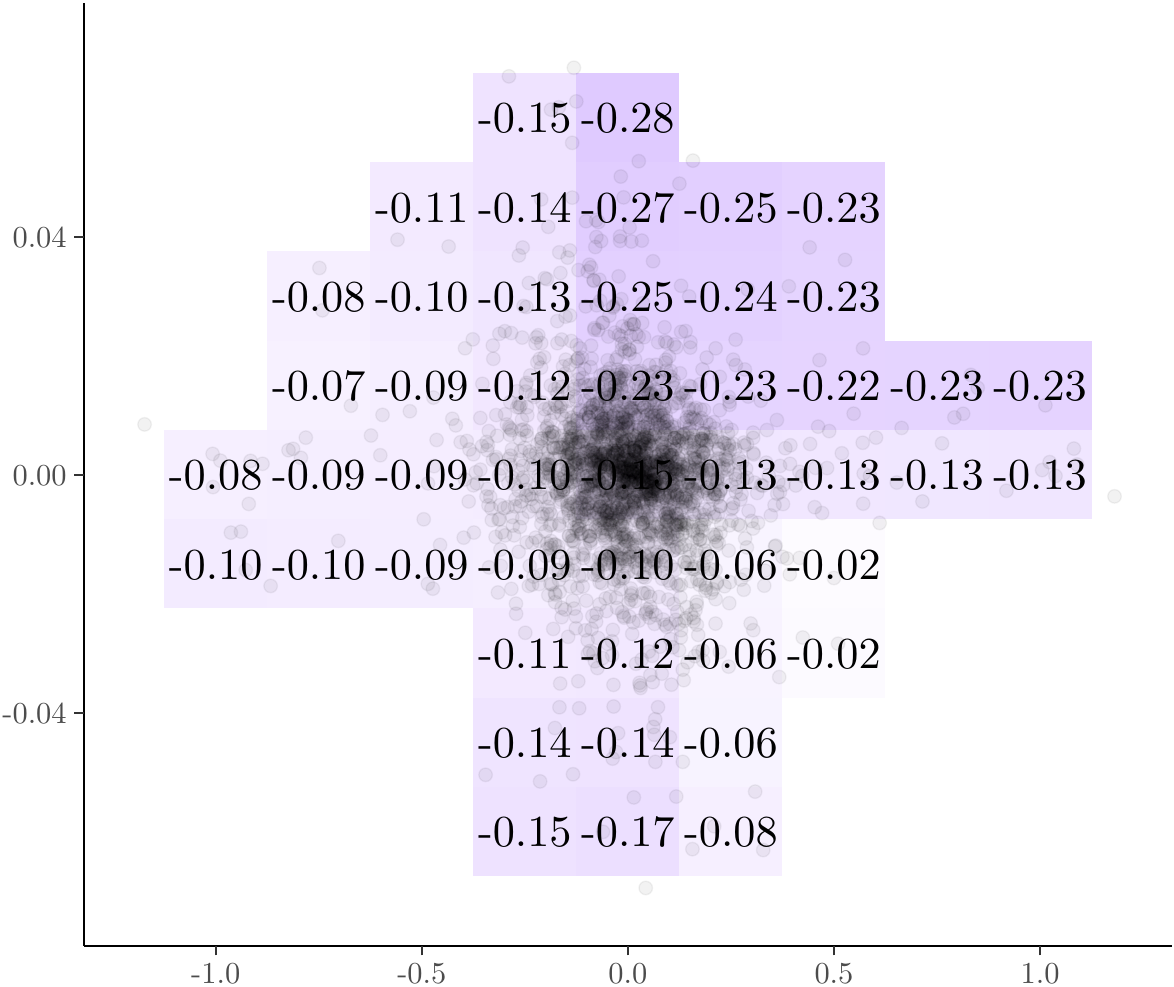} }\subfloat[Local partial dependence map of $R_t$ vs. $V_{t-1}^*$ on the \newline plane defined by $R_{t-1} = 0$\label{fig:realplots2}]{\includegraphics[width=0.49\linewidth]{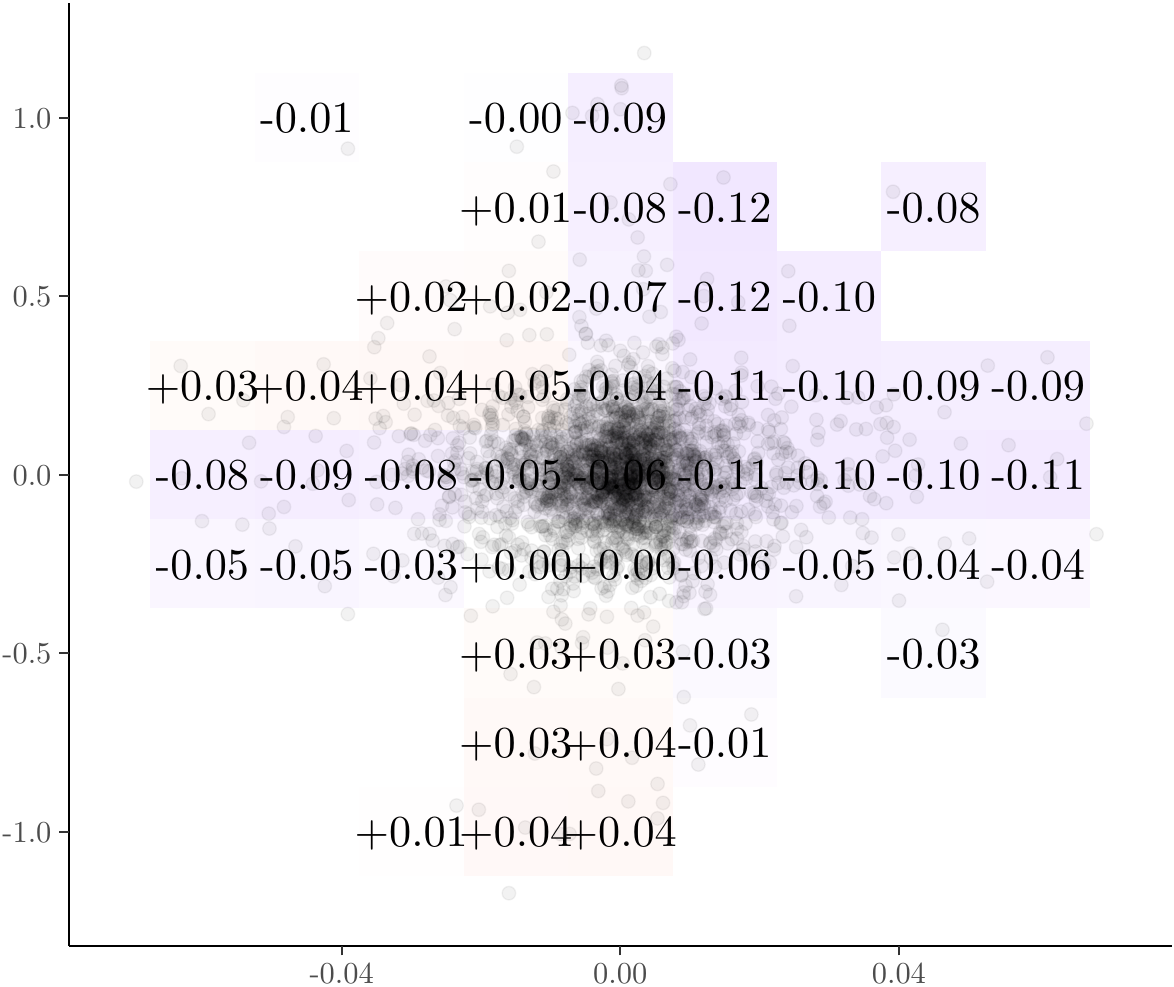} }\newline\subfloat[Local partial dependence map of $R_t$ vs. $V_{t-1}^*$ on the \newline plane defined by $R_{t-1} = -0.05$\label{fig:realplots3}]{\includegraphics[width=0.49\linewidth]{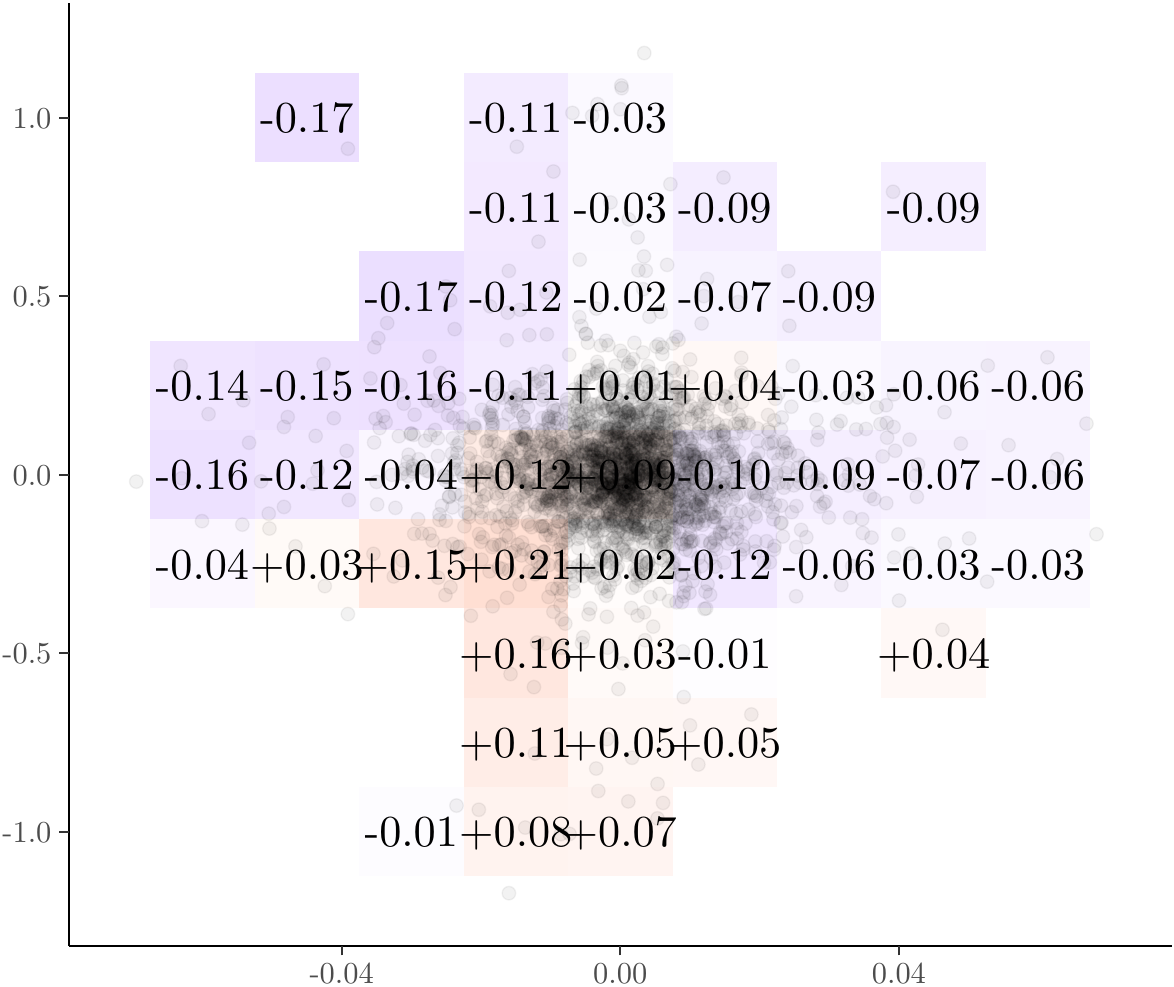} }\subfloat[Local partial dependence map of $R_t$ vs. $V_{t-1}^*$ on the \newline plane defined by $R_{t-1} = 0.05$\label{fig:realplots4}]{\includegraphics[width=0.49\linewidth]{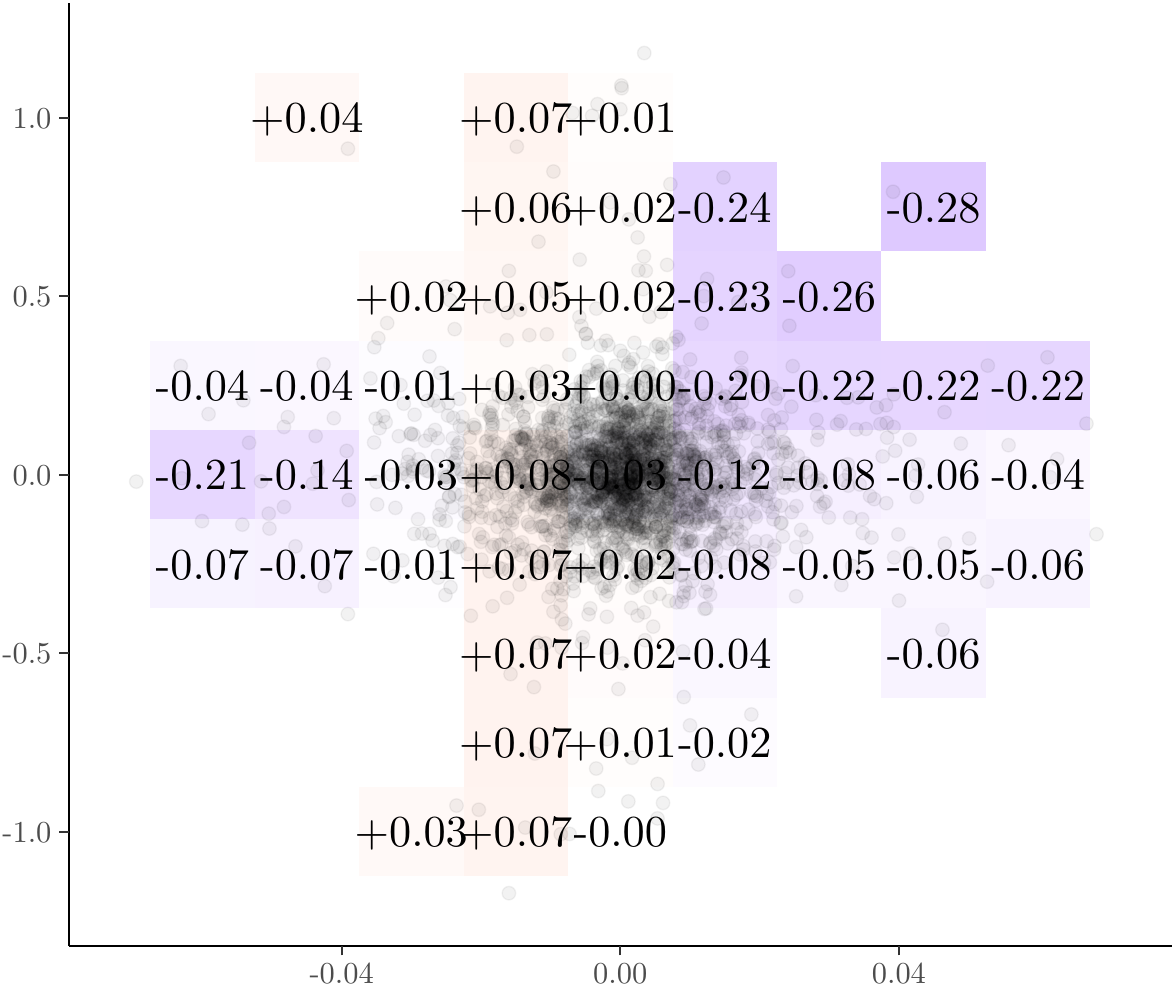} }\caption{Local partial dependence maps for the S\&P 500 data}\label{fig:realplots}
\end{figure}

Let us now, before we bring this section to its conclusion, briefly
demonstrate a practical implementation of the LGPC in a situation
involving real data. In particular, we will look at Granger causality,
which, it goes without saying, has been a central concept in economics
and econometrics ever since its inception by C. W. Granger
(\protect\hyperlink{ref-granger1969investigating}{1969}). In layman's
terms, the time series \(\{X_t\}\) Granger causes \(\{Y_t\}\) if past
and present values of \(\{X_t\}\) are helpful when predicting future
values of \(\{Y_t\}\). Formally, \(\{X_t\}\) Granger causes \(\{Y_t\}\)
if (C. W. Granger \protect\hyperlink{ref-granger1980testing}{1980})

\begin{equation}
Y_t \not\perp \mathcal{I}^*(t-1) \,\,|\,\, \mathcal{I}_{-X}^*(t-1),
\label{eq:granger}
\end{equation}

where \(\perp\) denotes independence, \(\not\perp\) denotes dependence,
and where \(\mathcal{I}^*(t-1)\) is all information available at time
\(t-1\) and \(\mathcal{I}_{-X}^*(t-1)\) is the same information, with
the exception of the values of \(\{X_t\}\) up to, but not including,
time \(t\). Of course, the hypothesis \eqref{eq:granger} can not, in
practice, be tested in its full generality. By taking only effects up to
the first lag into account, we may formulate a sufficient (but not
necessary) condition for \eqref{eq:granger}:
\(Y_t \not\perp X_{t-1} \,\,|\,\, Y_{t-1}\), the converse of which
constitutes a testable null hypothesis:

\begin{equation}
\textrm{H}_0: Y_t \perp X_{t-1} \,\,|\,\, Y_{t-1}.
\label{eq:grangernull}
\end{equation}

There are many ways to carry out this test in practice. The simplest is
based on the further restriction of linear relationships between
\(\{X_t\}\) and \(\{Y_t\}\) and is thus a test for \emph{conditional
uncorrelatedness} rather than conditional independence. Nonparametric
tests that have power against many nonlinear types of conditional
dependence have also been developed, and we refer to Section
\ref{chap:testing} for further references to this literature and a new
test for conditional independence based on the LGPC.

One fairly recent method for testing \eqref{eq:grangernull} is based on
the maximal conditional correlation and was introduced by Huang
(\protect\hyperlink{ref-huang2010testing}{2010}) and later extended to
the time series case by Cheng and Huang
(\protect\hyperlink{ref-cheng2012conditional}{2012}). The latter authors
use this test to examine whether trading volume Granger caused index
value, or vice versa, on the S\&P 500 stock index during the first
decade of this century. Using the daily price series \(\{P_t\}\) and
volume \(\{V_t\}\) they define the log-differenced series

\[R_t = 100\log\left(\frac{P_t}{P_{t-1}}\right) \,\,\, \textrm{ and } \,\,\, V_t^* = \log\left(\frac{V_t}{V_{t-1}}\right).\]
Further, they denote by \(R_{t-1} \not\Rightarrow V_t^*\) the Granger
non-causality from \(\{R_t\}\) to \(\{V_t^*\}\), which they analyze by
putting the following null hypothesis to the test:

\[\textrm{H}_0: V_t^* \perp R_{t-1} \,\, | \,\, V_{t-1}^*.\]

Granger non-causality in the opposite direction,
\(V_{t-1}^* \not\Rightarrow R_t\) is formulated and tested for in the
obvious way.

A linear test rejects \(R_{t-1} \not\Rightarrow V_t^*\), but not
\(V_{t-1}^* \not\Rightarrow R_t\). The nonparametric test by Cheng and
Huang (\protect\hyperlink{ref-cheng2012conditional}{2012}) rejects both,
leading to the natural conclusion that there are nonlinear relationships
in the latter situation that remain unseen by linear models. What is it
though, in this particular data set, that leads to such results?
Estimates of the local partial Gaussian correlation provide some clues
towards answering that question.

We obtain daily observations on price and trading volume on the S\&P 500
index from January 1st 2000 until December 31st 2009 on the S\&P 500
index from Yahoo Finance (\protect\hyperlink{ref-yahoo}{2018}), and plot
the estimated LGPC between \(V_t^*\) and \(R_{t-1}\) as defined in
Section \ref{chap:trivariate-full} on the plane defined by
\(V_{t-1}^* = 0\) in Figure \ref{fig:realplots1}. Departures from
\(\alpha(V_t^*, R_{t-1}) \equiv 0\) provide evidence against
\(R_{t-1} \not\Rightarrow V_{t}^*\). Similarly, we plot the estimated
LGPC between \(R_t\) and \(V^*_{t-1}\) on the plane \(R_{t-1} = 0\) in
Figure \ref{fig:realplots2} in which departures from
\(\alpha(R_t, V^*_{t-1}) \equiv 0\) provide evidence against
\(V^*_{t-1} \not\Rightarrow R_t\). We have used a proportionality
constant of \(c = 3.5\) to calculate the estimates (see discussion on
bandwidth selection in Section \ref{chap:simulated-examples}). The
observations can be seen in the background of both plots.

The differences between Figures \ref{fig:realplots1} and
\ref{fig:realplots2} are subtle, but important. In Figure
\ref{fig:realplots1}, \(\widehat\alpha(V_t^*, R_{t-1})\) is mostly
negative, especially in the data rich portions of the sample space
(other values of the conditioning variable \(V_{t-1}^*\) than zero give
similar pictures). Indeed, the \emph{global} partial correlation in this
situation is \(\widehat{\alpha}_{\textrm{glob}} = -0.086\), which is
significantly different from zero. The global partial correlation in the
second situation is very small in absolute value though:
\(\widehat{\alpha}_{\textrm{glob}} = -0.0018\), but Figure
\ref{fig:realplots2} still reveals departures from conditional
independence of similar magnitudes as in Figure \ref{fig:realplots1}.
The difference is that the estimated LGPC is positive (primarily in the
second and fourth quadrants) as well as negative (in the first and third
quadrants), but this pattern collapses to the constant global value
\(\alpha(R_t, V_{t-1}^*) = \alpha_{\textrm{glob}} \approx 0\) as the
bandwidths tend to infinity. In Figures \ref{fig:realplots3} and
\ref{fig:realplots4} we can explore the conditional dependence between
\(R_t\) and \(V_{t-1}\) at high and low levels of \(R_{t-1}\),
respectively, and we see even more clear differences in this dimension,
especially in the first and second quadrants.

The \(p\)-values for tests of the hypotheses \eqref{eq:grangernull} and
its opposite counterpart, using our new test for conditional
independence as presented in the next section, are both equal to 0.

\section{Testing for conditional independence}\label{chap:testing}

\subsection{The recent fauna of nonparametric tests}\label{fauna}

Property 3 in Section \ref{properties} states that the LGPC
characterizes conditional dependence within a large class of
distributions: \(X_1\) and \(X_2\) are independent given \(\X_3\) if and
only if the locally Gaussian partial correlation between them is
identically equal to zero everywhere on the sample space of
\((X_1, X_2, \X_3)\). The road is therefore short to a test for
conditional independence that may have power against a great deal of
nonlinear alternatives compared to a test based on the ordinary partial
correlation coefficient. One can, however, no longer claim that «few
methods exist for testing for conditional independence between \(X_1\)
and \(X_2\) given \(\X_3\) in a general nonparametric setting» as some
of the earlier references mentioned below do. In fact, the last decade
or so has seen the publication of many new tests for conditional
independence, some of which are presented along with rigorous derivation
of asymptotic properties and thorough simulations.

Su and White have published a series of such tests: Su and White
(\protect\hyperlink{ref-su2007consistent}{2007}) is based on detecting
differences between estimated characteristic functions (which is also
the method used by Wang and Hong
(\protect\hyperlink{ref-wang2017characteristic}{2017})), Su and White
(\protect\hyperlink{ref-su2008nonparametric}{2008}) is based on
estimating the Hellinger distance between conditional density estimates,
Su and White (\protect\hyperlink{ref-su2012conditional}{2012}) use local
polynomial quantile regression to test for conditional independence, and
Su and White (\protect\hyperlink{ref-su2014testing}{2014}) present a
test based on empirical likelihood. Huang
(\protect\hyperlink{ref-huang2010testing}{2010}) introduces the maximal
nonlinear conditional correlation which is used in a test for
conditional independence, in turn extended to dependent data by Cheng
and Huang (\protect\hyperlink{ref-cheng2012conditional}{2012}). Song
(\protect\hyperlink{ref-song2009testing}{2009}) constructs a test via
Rosenblatt transformations, while Bergsma
(\protect\hyperlink{ref-bergsma2011nonparametric}{2010}) and Bouezmarni,
Rombouts, and Taamouti
(\protect\hyperlink{ref-bouezmarni2012nonparametric}{2012}) present new
tests for conditional independence based on copula constructions.
Bouezmarni and Taamouti (\protect\hyperlink{ref-boue:taam:2014}{2014})
test for conditional independence by measuring the \(L_2\) distance
between estimated conditional distribution functions, and Patra, Sen,
and Székely (\protect\hyperlink{ref-patra2016on}{2016}) use empirical
transformations to translate conditional independence to joint
independence, tests for which exist in abundance in a large body of
literature stretching decades back in time. Wang et al.
(\protect\hyperlink{ref-wang2015conditional}{2015}) introduce the
\emph{conditional distance correlation} which they use to construct a
test for conditional independence. Most of the papers mentioned here
refer to a discussion paper by Linton and Gozalo
(\protect\hyperlink{ref-linton1997conditional}{1997}), who formulate
conditional independence in terms of probability statements which then
forms the basis of a test. A version of this work has been published
later as Linton and Gozalo
(\protect\hyperlink{ref-linton2014testing}{2014}). Also, Delgado and
Manteiga (\protect\hyperlink{ref-delgado2001significance}{2001}) develop
a test for conditional independence in a nonparametric regression
framework. Finally, we mention that there is a small literature on
testing by way of characterizing conditional independence via
reproducing kernel Hilbert spaces (RKHS), see, for example, Zhang et al.
(\protect\hyperlink{ref-zhang2012kernel}{2012}) and references therein.

Apart from some notational similarities with Zhang et al.
(\protect\hyperlink{ref-zhang2012kernel}{2012}), a test based on the
LGPC is quite different from the methods quoted above. First of all it
is semi-parametric and does not rely on traditional kernel estimates of
density-, distribution-, or characteristic functions. This opens up the
possibility for better power properties. Furthermore, due to our
transformation of the data \eqref{eq:trans} to marginal standard
normality, we do not necessarily have to specify weight functions in our
test functional to lessen the impact of outliers - unless of course we
wish to test for conditional independence in a specific portion of the
sample space.

On the other hand, we have to pay a price when imposing structure on the
dependence. Certain types of conditional dependence will remain
invisible to our test statistic, but simulation experiments show that
our test performs on par with, and sometimes better than existing, fully
nonparametric tests.

\subsection{A test for conditional independence based on the
LGPC}\label{thetest}

In accordance with many central references mentioned above, and indeed
with a larger literature on general independence testing based on local
measures of dependence, we construct a test statistic for testing

\[\textrm{H}_0: X_1 \perp X_2 \,\, | \,\, \X_3\] or, equivalently in
terms of the marginally Gaussian pseudo observations

\begin{equation}
\textrm{H}_0: Z_1 \perp Z_2 \,\, | \,\, \Z_3
\label{eq:ci-test}
\end{equation}

by aggregating our local measure of dependence over the sample space of
\(\X\) (or \(\Z\)). A natural test statistic on the \(z\)-scale is

\begin{equation}
T_{n,b} = \int_S h\left( \widehat \alpha_b(\z)  \right) \di F_n(\z),
\label{eq:teststatistic}
\end{equation}

where \(h(\cdot)\) is a real-valued function that is typically even and
non-negative for most standard applications, but not necessarily so, and
\(S \subseteq \mathbb{R}^p\) is an integration area that can be altered
in order to focus the test to specific portions of the sample space.
Standard laws of large numbers ensure that, under regularity conditions,
\(T_{n,b}\) converges in probability towards its population value
\[T = \int_S h\left(\alpha(\z)\right) \di F(\z).\]

This way, departures from many types of conditional independence lead to
large values of \(T_{n,b}\) that, if larger than a critical value
depending on the chosen significance level, leads to the rejection of
\eqref{eq:ci-test}. One might expect that approximate \(p\)-values for our
test can be readily extracted from the limiting distribution of the test
statistic \(T_{n,b}\), which can be derived along the same lines as
Lacal and Tjøstheim (\protect\hyperlink{ref-lacal2018estimating}{2018}).
Several authors, for example Teräsvirta, Tjøstheim, and Granger
(\protect\hyperlink{ref-terasvirta2010modelling}{2010}), have noted,
however, that asymptotic analysis of nonparametric test statistics on
the form \eqref{eq:teststatistic} tend to be too crude for finite sample
applications. We have therefore chosen to approximate the null
distribution of \(T_{n,b}\) using the bootstrap. The validity of the
bootstrap in such contexts has been examined by Lacal and Tjøstheim
(\protect\hyperlink{ref-lacal2018estimating}{2018}).

In accordance with our treatment so far, let
\(\{Z_{1,i}, Z_{2, i}, \Z_{3, i}\}\), \(i = 1,\ldots,n\) be observations
on the \(p\)-variate stochastic vector \(\Z\), \(p\geq3\), with \(Z_1\)
and \(Z_2\) being scalar and \(\Z_3\) being \(p-2\)-variate. In order to
calculate the statistic \(T_{n,b}\) for testing the null hypothesis
\(Z_1 \perp Z_2 \,\,|\,\, \Z_3\), we must estimate the joint local
correlation matrix \(\Sigma(\z)\) of \(\Z = (Z_1, Z_2, \Z_3)\), which we
may also use to estimate the conditional probability density functions
of \(Z_1|\Z_3\) and \(Z_2|\Z_3\) according to Otneim and Tjøstheim
(\protect\hyperlink{ref-otneim2017conditional}{2018}). We exploit this
in the following algorithm designed to produce approximate resampled
versions of \(T_{n,b}\):

\begin{enumerate}
\def\labelenumi{\arabic{enumi}.}
\tightlist
\item
  Use the local correlation estimates to estimate the conditional
  densities \(f_{Z_1|\Z_3}(\cdot)\) and \(f_{Z_2|\Z_3}(\cdot)\) by means
  of the method by Otneim and Tjøstheim
  (\protect\hyperlink{ref-otneim2017conditional}{2018}). In practice, in
  order to reduce the computational load, we evaluate
  \(\widehat f_{Z_1|\Z_3}\) and \(\widehat f_{Z_2|\Z_3}\) on a fine grid
  on their support, over which a continuous representation of the
  estimates are produced using cubic splines.
\item
  Using the accept-reject algorithm, generate \(B\) samples, each of
  size \(n\), from \(\widehat f_{Z_1|\Z_3}\) and
  \(\widehat f_{Z_2|\Z_3}\), leading to \(B\) replicates
  \[Z_m^* = \Big\{\{Z_{1,i}^*, Z_{2,i}^*, \Z_{3,i}\}_i\Big\}_m, \qquad i = 1,\ldots,n, \,\, m = 1,\ldots,B,\]
  of \(\Z\) under the null hypothesis.
\item
  Calculate \(\{T_{n,b}\}_m^*\), \(m = 1,\ldots,B\) for the replicated
  data sets and obtain the approximate \(p\)-value for the conditional
  independence test.
\end{enumerate}

\subsection{Comparing with other tests}\label{chap:testexamples}

Su and White (\protect\hyperlink{ref-su2008nonparametric}{2008})
formulate a simulation experiment for evaluating their nonparametric
test for conditional independence by generating data from 10 different
data generating processes (DGPs), and then check the power and level for
their test. Many of the later works that were discussed in Section
\ref{fauna} contain very similar experiments, and some of them even
present simulation results for tests published prior to Su and White
(\protect\hyperlink{ref-su2008nonparametric}{2008}). This allows us to
present comprehensive comparisons between the new test presented above,
and many competitors.

Let \((\epsilon_{1,t}, \epsilon_{2,t}, \epsilon_{3,t})\) be IID
observations from a \(\mathcal{N}(0, I_3)\)-distribution, where \(I_3\)
is the \(3\times3\) identity matrix. We will test
H\(_0: X_{1,t} \perp X_{2,t} \,\,|\,\, X_{3,t}\) in the following 10
cases taken from Su and White
(\protect\hyperlink{ref-su2008nonparametric}{2008}), which cover various
types of linear and nonlinear time series dependence:

\begin{enumerate}
\def\labelenumi{\arabic{enumi}.}
\tightlist
\item
  \((X_{1,t}, X_{2,t}, X_{3,t}) = (\epsilon_{1,t}, \epsilon_{2,t}, \epsilon_{3,t})\).
\item
  \(X_{1,t} = 0.5X_{1, t-1} + \epsilon_{1,t}\),
  \(X_{2,t} = 0.5X_{2,t-1} + \epsilon_{2,t}\), \(X_{3,t} = X_{1,t-1}\).
\item
  \(X_{1,t} = \epsilon_{1,t}\sqrt{0.01 + 0.5X_ {t-1}^2}\),
  \(X_{2,t} = 0.5X_{2, t-1} + \epsilon_{2,t}\), \(X_{3,t} = X_{1,t-1}\).
\item
  \(X_{1,t} = \epsilon_{1,t}\sqrt{h_{1,t}}\),
  \(X_{2,t} = \epsilon_{2,t}\sqrt{h_{2,t}}\), \(X_{3,t} = X_{1,t-1}\),
  \(h_{1,t} = 0.01 + 0.9h_{1,t-1} + 0.05X_{1,t-1}^2\),
  \(h_{2,t} = 0.01 + 0.9h_{2,t-1} + 0.05X_{2,t-1}^2\).
\item
  \(X_{1,t} = 0.5X_{1,t-1} + 0.5X_{2,t} + \epsilon_{1,t}\),
  \(X_{2,t} = 0.5X_{2,t-1} + \epsilon_{2,t}\), \(X_{3,t} = X_{1,t-1}\).
\item
  \(X_{1,t} = 0.5X_{1,t-1} + 0.5X_{2,t}^2 + \epsilon_{1,t}\),
  \(X_{2,t} = 0.5X_{2,t-1} + \epsilon_{2,t}\), \(X_{3,t} = X_{1,t-1}\).
\item
  \(X_{1,t} = 0.5X_{1,t-1}0.5X_{2,t} + \epsilon_{1,t}\),
  \(X_{2,t} = 0.5X_{2,t-1} + \epsilon_{2,t}\), \(X_{3,t} = X_{1,t-1}\).
\item
  \(X_{1,t} = 0.5X_{1,t-1} + 0.5X_{2,t}\epsilon_{1,t}\),
  \(X_{2,t} = 0.5X_{2,t-1} + \epsilon_{2,t}\), \(X_{3,t} = X_{1,t-1}\).
\item
  \(X_{1,t} = \epsilon_{1,t}\sqrt{0.01 + 0.5X_{1,t-1}^2 + 0.25X_{2,t}}\),
  \(X_{2,t} = X_{2,t-1} + \epsilon_{2,t}\), \(X_{3,t} = X_{1,t-1}\).
\item
  \(X_{1,t} = \epsilon_{1,t}\sqrt{h_{1,t}}\),
  \(X_{2,t} = \epsilon_{2,t}\sqrt{h_{2,t}}\), \(X_{3,t} = X_{1,t-1}\),
  \(h_{1,t} = 0.01 + 0.1h_{1, t-1} + 0.4X_{1,t-1}^2 + 0.5X_{2,t}^2\),
  \(h_{2,t} = 0.01 + 0.9h_{2, t-1} + 0.5Y_t^2\).
\end{enumerate}

The null hypotheses of conditional independence between \(X_{1,t}\) and
\(X_{2,t}\) given \(X_{3,t}\) is true for DGPs 1--4, and these will be
used to check the level of the test, while for DGPs 5--10 we measure the
power. By evaluating our test using the sample sizes \(n = 100\) and
\(n = 200\) at the \(5\%\) level, we can harvest a great number of
corresponding results from the literature, and they are presented in
tables 1 and 2.

The first set of simulation results are collected from Su and White
(\protect\hyperlink{ref-su2008nonparametric}{2008}): LIN is a standard
linear Granger causality test, CM and KS are tests by Linton and Gozalo
(\protect\hyperlink{ref-linton1997conditional}{1997}) that use
statistics of the two versions of the nonparametric test developed by
Delgado and Manteiga
(\protect\hyperlink{ref-delgado2001significance}{2001}). Finally, HEL
refers to the Hellinger distance test that Su and White
(\protect\hyperlink{ref-su2008nonparametric}{2008}) present, at
different levels of smoothing. Their constant \(c\) has the same meaning
as our smoothing constant discussed in Section \ref{chap:examples}.

\renewcommand{\arraystretch}{1.2}
\begin{table}[t]
\begin{tabular}{l|rrrr|rrrrrr}
\toprule
&\multicolumn{4}{c}{Level} & \multicolumn{6}{c}{Power} \\
$\downarrow$ Test $\mid$ DGP $\rightarrow$ & 1 & 2 & 3 & 4 & 5 & 6 & 7 & 8 & 9 & 10 \\
\midrule
CHF&\texttt{0.034}&\texttt{0.058}&\texttt{-}&\texttt{-}&\texttt{0.780}&\texttt{0.792}&\texttt{0.520}&\texttt{0.780}&\texttt{0.728}&\texttt{0.580}\\
CM&\texttt{0.054}&\texttt{0.058}&\texttt{0.060}&\texttt{0.048}&\texttt{0.920}&\texttt{0.548}&\texttt{0.504}&\texttt{0.412}&\texttt{0.384}&\texttt{0.188}\\
HEL, $\scriptstyle c = 1$&\texttt{0.096}&\texttt{0.060}&\texttt{0.048}&\texttt{0.072}&\texttt{0.668}&\texttt{0.756}&\texttt{0.388}&\texttt{0.860}&\texttt{0.828}&\texttt{0.680}\\
HEL, $\scriptstyle c = 1.5$&\texttt{0.068}&\texttt{0.056}&\texttt{0.052}&\texttt{0.056}&\texttt{0.888}&\texttt{0.940}&\texttt{0.512}&\texttt{0.924}&\texttt{0.952}&\texttt{0.812}\\
HEL, $\scriptstyle c = 2$&\texttt{0.072}&\texttt{0.036}&\texttt{0.072}&\texttt{0.048}&\texttt{0.952}&\texttt{\underline{0.944}}&\texttt{0.576}&\texttt{0.940}&\texttt{\underline{0.988}}&\texttt{\underline{0.912}}\\
KS&\texttt{0.042}&\texttt{0.056}&\texttt{0.056}&\texttt{0.040}&\texttt{0.780}&\texttt{0.404}&\texttt{0.380}&\texttt{0.288}&\texttt{0.292}&\texttt{0.156}\\
\rowcolor{Gray}LGPC, $\scriptstyle c = 1.0$&\texttt{0.054}&\texttt{0.048}&\texttt{0.046}&\texttt{0.046}&\texttt{0.910}&\texttt{0.722}&\texttt{0.559}&\texttt{\underline{0.990}}&\texttt{0.968}&\texttt{0.866}\\
\rowcolor{Gray}LGPC, $\scriptstyle c = 1.4$&\texttt{0.047}&\texttt{0.043}&\texttt{0.046}&\texttt{0.047}&\texttt{0.971}&\texttt{0.855}&\texttt{0.727}&\texttt{0.969}&\texttt{0.916}&\texttt{0.765}\\
LIN&\texttt{0.044}&\texttt{0.061}&\texttt{0.050}&\texttt{0.060}&\texttt{\underline{0.999}}&\texttt{0.337}&\texttt{0.213}&\texttt{0.126}&\texttt{0.163}&\texttt{0.153}\\
MCC, $\scriptstyle c = 1$&\texttt{0.046}&\texttt{0.050}&\texttt{0.050}&\texttt{0.047}&\texttt{0.746}&\texttt{0.717}&\texttt{0.400}&\texttt{0.873}&\texttt{0.566}&\texttt{0.320}\\
MCC, $\scriptstyle c = 1.5$&\texttt{0.040}&\texttt{0.052}&\texttt{0.056}&\texttt{0.055}&\texttt{0.814}&\texttt{0.779}&\texttt{0.329}&\texttt{0.889}&\texttt{0.618}&\texttt{0.341}\\
MCC, $\scriptstyle c = 2$&\texttt{0.041}&\texttt{0.050}&\texttt{0.053}&\texttt{0.062}&\texttt{0.852}&\texttt{0.793}&\texttt{0.218}&\texttt{0.860}&\texttt{0.631}&\texttt{0.348}\\
SCM&\texttt{0.076}&\texttt{0.060}&\texttt{0.084}&\texttt{0.064}&\texttt{0.924}&\texttt{0.464}&\texttt{0.352}&\texttt{0.500}&\texttt{0.224}&\texttt{0.196}\\
SEL&\texttt{0.054}&\texttt{0.038}&\texttt{-}&\texttt{-}&\texttt{0.840}&\texttt{0.856}&\texttt{\underline{0.760}}&\texttt{0.904}&\texttt{0.716}&\texttt{0.556}\\
SKS&\texttt{0.064}&\texttt{0.056}&\texttt{0.088}&\texttt{0.068}&\texttt{0.728}&\texttt{0.236}&\texttt{0.288}&\texttt{0.340}&\texttt{0.120}&\texttt{0.112}\\
\bottomrule
\end{tabular}
\caption{Level and power, $n = 100$}
\label{tab:n100}
\end{table}

Next, we move to Su and White
(\protect\hyperlink{ref-su2014testing}{2014}), who provide simulations
for their test for conditional independence based on the empirical
likelihood (SEL), as well as the test by Su and White
(\protect\hyperlink{ref-su2007consistent}{2007}) that is based on
properties of the conditional characteristic function (CHF). Cheng and
Huang (\protect\hyperlink{ref-cheng2012conditional}{2012}) provide
simulations for their tests based on the maximal conditional correlation
(MCC) at various levels of smoothing.

\renewcommand{\arraystretch}{1.2}
\begin{table}[t]
\begin{tabular}{l|rrrr|rrrrrr}
\toprule
&\multicolumn{4}{c}{Level} & \multicolumn{6}{c}{Power} \\
$\downarrow$ Test $\mid$ DGP $\rightarrow$ & 1 & 2 & 3 & 4 & 5 & 6 & 7 & 8 & 9 & 10 \\
\midrule
BRT, $\scriptstyle c = 1$&\texttt{0.037}&\texttt{0.025}&\texttt{0.028}&\texttt{0.032}&\texttt{0.995}&\texttt{0.996}&\texttt{\underline{0.979}}&\texttt{0.989}&\texttt{0.904}&\texttt{0.785}\\
BRT, $\scriptstyle c = 1.5$&\texttt{0.044}&\texttt{0.025}&\texttt{0.025}&\texttt{0.037}&\texttt{0.971}&\texttt{0.993}&\texttt{0.931}&\texttt{0.997}&\texttt{0.931}&\texttt{0.759}\\
BRT, $\scriptstyle c = 2$&\texttt{0.064}&\texttt{0.023}&\texttt{0.023}&\texttt{0.052}&\texttt{0.943}&\texttt{0.979}&\texttt{0.873}&\texttt{0.997}&\texttt{0.912}&\texttt{0.728}\\
BT, $\scriptstyle c_1 = 1, c_2 = 1$&\texttt{0.047}&\texttt{0.051}&\texttt{0.041}&\texttt{0.053}&\texttt{0.996}&\texttt{0.812}&\texttt{0.852}&\texttt{\underline{1.000}}&\texttt{0.936}&\texttt{-}\\
BT, $\scriptstyle c_1 = 0.85, c_2 = 0.7$&\texttt{0.048}&\texttt{0.044}&\texttt{0.064}&\texttt{0.056}&\texttt{0.988}&\texttt{0.728}&\texttt{0.792}&\texttt{\underline{1.000}}&\texttt{0.908}&\texttt{-}\\
BT, $\scriptstyle c_1 = 0.75, c_2 = 0.6$&\texttt{0.036}&\texttt{0.048}&\texttt{0.052}&\texttt{0.052}&\texttt{0.976}&\texttt{0.719}&\texttt{0.808}&\texttt{\underline{1.000}}&\texttt{0.896}&\texttt{-}\\
CHF&\texttt{0.046}&\texttt{0.042}&\texttt{-}&\texttt{-}&\texttt{0.976}&\texttt{0.988}&\texttt{0.820}&\texttt{0.952}&\texttt{0.944}&\texttt{0.864}\\
CM&\texttt{0.044}&\texttt{0.056}&\texttt{0.060}&\texttt{0.048}&\texttt{0.992}&\texttt{0.740}&\texttt{0.788}&\texttt{0.680}&\texttt{0.476}&\texttt{0.360}\\
HEL, $\scriptstyle c = 1$&\texttt{0.064}&\texttt{0.052}&\texttt{0.080}&\texttt{0.080}&\texttt{0.900}&\texttt{0.960}&\texttt{0.596}&\texttt{0.992}&\texttt{0.968}&\texttt{0.880}\\
HEL, $\scriptstyle c = 1.5$&\texttt{0.064}&\texttt{0.056}&\texttt{0.048}&\texttt{0.036}&\texttt{0.980}&\texttt{\underline{1.000}}&\texttt{0.808}&\texttt{0.992}&\texttt{0.972}&\texttt{0.972}\\
HEL, $\scriptstyle c = 2$&\texttt{0.044}&\texttt{0.060}&\texttt{0.056}&\texttt{0.048}&\texttt{\underline{1.000}}&\texttt{\underline{1.000}}&\texttt{0.864}&\texttt{\underline{1.000}}&\texttt{\underline{1.000}}&\texttt{\underline{0.996}}\\
KS&\texttt{0.068}&\texttt{0.053}&\texttt{0.048}&\texttt{0.084}&\texttt{0.952}&\texttt{0.552}&\texttt{0.660}&\texttt{0.532}&\texttt{0.336}&\texttt{0.284}\\
\rowcolor{Gray}LGPC, $\scriptstyle c = 1.0$&\texttt{0.039}&\texttt{0.052}&\texttt{0.054}&\texttt{0.054}&\texttt{0.995}&\texttt{0.948}&\texttt{0.818}&\texttt{\underline{1.000}}&\texttt{\underline{1.000}}&\texttt{0.985}\\
\rowcolor{Gray}LGPC, $\scriptstyle c = 1.4$&\texttt{0.042}&\texttt{0.057}&\texttt{0.058}&\texttt{0.042}&\texttt{\underline{1.000}}&\texttt{0.993}&\texttt{0.956}&\texttt{\underline{1.000}}&\texttt{\underline{1.000}}&\texttt{0.958}\\
LIN&\texttt{0.043}&\texttt{0.053}&\texttt{0.042}&\texttt{0.050}&\texttt{\underline{1.000}}&\texttt{0.354}&\texttt{0.250}&\texttt{0.113}&\texttt{0.172}&\texttt{0.143}\\
MCC, $\scriptstyle c = 1$&\texttt{0.049}&\texttt{0.051}&\texttt{0.057}&\texttt{0.054}&\texttt{0.982}&\texttt{0.983}&\texttt{0.831}&\texttt{\underline{1.000}}&\texttt{0.947}&\texttt{0.679}\\
MCC, $\scriptstyle c = 1.5$&\texttt{0.046}&\texttt{0.048}&\texttt{0.049}&\texttt{0.053}&\texttt{0.995}&\texttt{0.989}&\texttt{0.872}&\texttt{\underline{1.000}}&\texttt{0.968}&\texttt{0.738}\\
MCC, $\scriptstyle c = 2$&\texttt{0.045}&\texttt{0.045}&\texttt{0.047}&\texttt{0.057}&\texttt{0.997}&\texttt{0.995}&\texttt{0.735}&\texttt{\underline{1.000}}&\texttt{0.971}&\texttt{0.745}\\
SCM&\texttt{0.048}&\texttt{0.060}&\texttt{0.064}&\texttt{0.068}&\texttt{0.980}&\texttt{0.648}&\texttt{0.620}&\texttt{0.720}&\texttt{0.352}&\texttt{0.280}\\
SEL&\texttt{0.052}&\texttt{0.033}&\texttt{-}&\texttt{-}&\texttt{0.992}&\texttt{\underline{1.000}}&\texttt{0.972}&\texttt{\underline{1.000}}&\texttt{0.884}&\texttt{0.864}\\
SKS&\texttt{0.056}&\texttt{0.028}&\texttt{0.064}&\texttt{0.072}&\texttt{0.964}&\texttt{0.324}&\texttt{0.512}&\texttt{0.552}&\texttt{0.148}&\texttt{0.136}\\
\bottomrule
\end{tabular}
\caption{Level and power, $n = 200$}
\label{tab:n200}
\end{table}

Finally, we include results from simulations using our new test based on
the LGPC, using the trivariate specification defined in Section
\ref{chap:trivariate-full}, for two different levels of smoothing, and
include them in Tables \ref{tab:n100} and \ref{tab:n200}. We highlight
the new results in grey to indicate that they, as opposed to all the
other numbers, have not appeared in the literature before. Also, to the
best of our knowledge, these results have not been compared
simultaneously before.

We see in Table \ref{tab:n100} that our test has the correct level and
is quite powerful against all alternative specifications of conditional
dependence, in the additive models 5 and 6, as well as the remaining
examples 7 to 10, that are more multiplicative in nature.

When \(n=200\) we can include simulation results reported by Bouezmarni,
Rombouts, and Taamouti
(\protect\hyperlink{ref-bouezmarni2012nonparametric}{2012}) for their
conditional independence test based on measuring the Hellinger distance
between copula density estimates (shortened BRT from the names of the
authors) as well as the results reported by Bouezmarni and Taamouti
(\protect\hyperlink{ref-boue:taam:2014}{2014}) on their test based on
\(L_2\) distances between estimated conditional distribution functions,
which is abbreviated by BT. We include in Table \ref{tab:n200} all
results from these papers, which amounts to three levels of smoothing
for each method. We see that the test based on the LGPC, in particular
the case with \(c=1.4\), exhibits the best over-all performance among
the examples listed in the table, which, again to the best of our
knowledge, include all such simulation results that have been published
to date. It is seen that the standard linear Granger causality test
(LIN) has a miserable performance in these examples.

\renewcommand{\arraystretch}{1.2}
\begin{table}[t]
\begin{tabular}{ll|rr|rrrrrr}
\toprule
&&\multicolumn{2}{c}{Level} & \multicolumn{6}{c}{Power} \\
Sample size & $\downarrow$ Test $\mid$ DGP $\rightarrow$ & 1$^{\prime}$ & 2$^{\prime}$ & 5$^{\prime}$ & 6$^{\prime}$ & 7$^{\prime}$ & 8$^{\prime}$ & 9$^{\prime}$ & 10$^{\prime}$ \\
\midrule
&&&&&&&&& \\ 
$n = 100$ & CHF&\texttt{0.028}&\texttt{0.042}&\texttt{0.720}&\texttt{0.704}&\texttt{0.412}&\texttt{0.564}&\texttt{\underline{0.460}}&\texttt{\underline{0.556}}\\
&CM&\texttt{0.028}&\texttt{0.016}&\texttt{0.656}&\texttt{0.360}&\texttt{0.108}&\texttt{0.512}&\texttt{0.164}&\texttt{0.208}\\
&KS&\texttt{0.040}&\texttt{0.020}&\texttt{0.400}&\texttt{0.284}&\texttt{0.056}&\texttt{0.380}&\texttt{0.124}&\texttt{0.176}\\
\rowcolor{Gray}&LGPC, $\scriptstyle c = 1.75$&\texttt{0.048}&\texttt{0.033}&\texttt{\underline{0.951}}&\texttt{0.668}&\texttt{0.294}&\texttt{0.517}&\texttt{0.406}&\texttt{0.538}\\
&SEL&\texttt{0.052}&\texttt{0.040}&\texttt{0.844}&\texttt{\underline{0.828}}&\texttt{\underline{0.620}}&\texttt{\underline{0.568}}&\texttt{0.440}&\texttt{0.528}\\
&&&&&&&&& \\ 
$n = 200$ & CHF&\texttt{0.030}&\texttt{0.040}&\texttt{0.948}&\texttt{0.944}&\texttt{0.748}&\texttt{0.828}&\texttt{\underline{0.724}}&\texttt{\underline{0.860}}\\
&CM&\texttt{0.050}&\texttt{0.032}&\texttt{0.940}&\texttt{0.588}&\texttt{0.304}&\texttt{0.792}&\texttt{0.304}&\texttt{0.364}\\
&KS&\texttt{0.046}&\texttt{0.024}&\texttt{0.776}&\texttt{0.432}&\texttt{0.168}&\texttt{0.696}&\texttt{0.216}&\texttt{0.284}\\
\rowcolor{Gray}&LGPC, $\scriptstyle c = 1.75$&\texttt{0.059}&\texttt{0.035}&\texttt{\underline{0.998}}&\texttt{0.923}&\texttt{0.415}&\texttt{0.769}&\texttt{0.698}&\texttt{0.833}\\
&SEL&\texttt{0.058}&\texttt{0.026}&\texttt{0.972}&\texttt{\underline{0.988}}&\texttt{\underline{0.932}}&\texttt{\underline{0.832}}&\texttt{0.684}&\texttt{0.832}\\
&&&&&&&&& \\ 
$n = 400$ & CHF&\texttt{0.040}&\texttt{0.036}&\texttt{\underline{1.000}}&\texttt{0.984}&\texttt{0.972}&\texttt{0.996}&\texttt{0.920}&\texttt{\underline{0.984}}\\
&CM&\texttt{0.056}&\texttt{0.024}&\texttt{\underline{1.000}}&\texttt{0.884}&\texttt{0.552}&\texttt{0.980}&\texttt{0.556}&\texttt{0.668}\\
&KS&\texttt{0.060}&\texttt{0.024}&\texttt{\underline{1.000}}&\texttt{0.732}&\texttt{0.324}&\texttt{0.952}&\texttt{0.384}&\texttt{0.524}\\
\rowcolor{Gray}&LGPC, $\scriptstyle c = 1.75$&\texttt{0.048}&\texttt{0.018}&\texttt{\underline{1.000}}&\texttt{0.996}&\texttt{0.584}&\texttt{0.960}&\texttt{\underline{0.928}}&\texttt{0.978}\\
&SEL&\texttt{0.040}&\texttt{0.030}&\texttt{\underline{1.000}}&\texttt{\underline{1.000}}&\texttt{\underline{1.000}}&\texttt{\underline{1.000}}&\texttt{0.836}&\texttt{0.884}\\
\bottomrule
\end{tabular}
\caption{Level and power, 4-dimensional data}
\label{tab:dim4}
\end{table}

\renewcommand{\arraystretch}{1.2}
\begin{table}[t]
\begin{tabular}{ll|rr|rrrrrr}
\toprule
&&\multicolumn{2}{c}{Level} & \multicolumn{6}{c}{Power} \\
Sample size & $\downarrow$ Test $\mid$ DGP $\rightarrow$ & 1$^{\prime\prime}$ & 2$^{\prime\prime}$ & 5$^{\prime\prime}$ & 6$^{\prime\prime}$ & 7$^{\prime\prime}$ & 8$^{\prime\prime}$ & 9$^{\prime\prime}$ & 10$^{\prime\prime}$ \\
\midrule
&&&&&&&&& \\ 
\rowcolor{Gray}$n = 100$ & LGPC, $\scriptstyle c = 1.75$&\texttt{0.068}&\texttt{0.049}&\texttt{\underline{0.911}}&\texttt{\underline{0.567}}&\texttt{\underline{0.295}}&\texttt{\underline{0.266}}&\texttt{\underline{0.380}}&\texttt{\underline{0.521}}\\
&&&&&&&&& \\ 
$n = 200$ & CHF&\texttt{0.028}&\texttt{0.022}&\texttt{0.964}&\texttt{0.952}&\texttt{0.668}&\texttt{\underline{0.852}}&\texttt{\underline{0.552}}&\texttt{\underline{0.856}}\\
&CM&\texttt{0.050}&\texttt{0.028}&\texttt{0.756}&\texttt{0.484}&\texttt{0.192}&\texttt{0.788}&\texttt{0.260}&\texttt{0.448}\\
&KS&\texttt{0.048}&\texttt{0.032}&\texttt{0.500}&\texttt{0.380}&\texttt{0.096}&\texttt{0.660}&\texttt{0.212}&\texttt{0.372}\\
\rowcolor{Gray}&LGPC, $\scriptstyle c = 1.75$&\texttt{0.062}&\texttt{0.050}&\texttt{0.993}&\texttt{0.776}&\texttt{0.388}&\texttt{0.363}&\texttt{0.545}&\texttt{0.794}\\
&SEL&\texttt{0.056}&\texttt{0.026}&\texttt{\underline{0.996}}&\texttt{\underline{0.980}}&\texttt{\underline{0.860}}&\texttt{0.816}&\texttt{0.344}&\texttt{0.680}\\
&&&&&&&&& \\ 
$n = 400$ & CHF&\texttt{0.032}&\texttt{0.034}&\texttt{\underline{1.000}}&\texttt{0.972}&\texttt{0.928}&\texttt{0.884}&\texttt{\underline{0.792}}&\texttt{\underline{0.972}}\\
&CM&\texttt{0.050}&\texttt{0.032}&\texttt{0.992}&\texttt{0.728}&\texttt{0.400}&\texttt{\underline{0.912}}&\texttt{0.390}&\texttt{0.620}\\
&KS&\texttt{0.044}&\texttt{0.036}&\texttt{0.840}&\texttt{0.552}&\texttt{0.220}&\texttt{0.880}&\texttt{0.306}&\texttt{0.568}\\
\rowcolor{Gray}&LGPC, $\scriptstyle c = 1.75$&\texttt{0.048}&\texttt{0.030}&\texttt{\underline{1.000}}&\texttt{0.938}&\texttt{0.519}&\texttt{0.538}&\texttt{0.740}&\texttt{0.956}\\
&SEL&\texttt{0.056}&\texttt{0.038}&\texttt{\underline{1.000}}&\texttt{\underline{1.000}}&\texttt{\underline{1.000}}&\texttt{0.888}&\texttt{0.616}&\texttt{0.876}\\
\bottomrule
\end{tabular}
\caption{Level and power, 5-dimensional data}
\label{tab:dim5}
\end{table}

Su and White (\protect\hyperlink{ref-su2014testing}{2014}) then define
two extensions to a subset of the data generating processes defined
above. The first extension turns the conditioning variable in DGP1-DGP2
and DGP5-DGP10 \(X_{3,t}\) into a bivariate vector \(\X_{3,t}\), where
we define DGP1\(^{\prime}\) in the same way as above,

\begin{itemize}
\item[1$^{\prime}$.] $(X_{1,t}, X_{2,t}, \X_{3,t}) = (\epsilon_{1,t}, \epsilon_{2,t}, \fepsilon_{3,t})$,
\end{itemize}

but where \(\fepsilon_{3,t} \sim \mathcal{N}(0, \bm{I}_2)\), and where
we define DGP2\(^{\prime}\) and DGP5\(^{\prime}\)-DGP10\(^{\prime}\) by
setting \(\X_{3,t} = (X_{1, t-1}, X_{1, t-2})\), keeping \(X_{2,t}\) as
above, and,

\begin{itemize}
\tightlist
\item[2$^{\prime}$.] $X_{1,t} = 0.5X_{1,t-1} + 0.25X_{1, t-2} + \epsilon_{1,t}$,
\item[5$^{\prime}$.] $X_{1,t} = 0.5X_{1,t-1} + 0.25X_{1, t-2} + 0.5X_{2,t} + \epsilon_{1,t}$,
\item[6$^{\prime}$.] $X_{1,t} = 0.5X_{1,t-1} + 0.25X_{1, t-2} + 0.5X_{2,t}^2 + \epsilon_{1,t}$,
\item[7$^{\prime}$.] $X_{1,t} = 0.5X_{1,t-1}X_{2, t} + 0.25X_{1,t-2}+ \epsilon_{1,t}$,
\item[8$^{\prime}$.] $X_{1,t} = 0.5X_{1,t-1} + 0.25X_{1,t-2} + 0.5X_{2, t}\epsilon_{1,t}$,
\item[9$^{\prime}$.] $X_{1,t} = \sqrt{h_t}\epsilon_{1,t}$, $h_t = 0.01 + 0.5X_{1,t-1}^2 + 0.25X_{1,t-2}^2 + 0.25X_{2,t}^2$, and
\item[10$^{\prime}$.] Same as DGP10 above, except for the new definition of $\X_{3,t}$.
\end{itemize}

We report the level and power results obtained by Su and White
(\protect\hyperlink{ref-su2014testing}{2014}) for the methods CHF, CM,
KS and SEL on these data by testing the null hypothesis
\[\textrm{H}_0: X_{t,1} \perp X_{t,2} \,\, | \,\, \X_{t,3},\] and
include results from our new test based on the multivariate
simplification of the LGPC as defined in Section \ref{chap:bivariate} in
table \ref{tab:dim4}. The results are quite acceptable and compares well
with other nonparametric methods.

The second extension introduced by Su and White
(\protect\hyperlink{ref-su2014testing}{2014}) increases the dimension of
the conditioning variable \(\X_{3,t}\) once more, so that
\(\fepsilon_{3,t} \sim \mathcal{N}(0, \bm{I}_3)\), and where
DGP2\(^{\prime\prime}\) and
DGP5\(^{\prime\prime}\)-DGP10\(^{\prime\prime}\) are defined by by
setting \(\X_{3,t} = (X_{1, t-1}, X_{1, t-2}, X_{1, t-3})\), and,

\begin{itemize}
\tightlist
\item[2$^{\prime\prime}$.] $X_{1,t} = 0.5X_{1,t-1} + 0.25X_{1, t-2} + 0.125X_{1,t-3} + \epsilon_{1,t}$,
\item[5$^{\prime\prime}$.] $X_{1,t} = 0.5X_{1,t-1} + 0.25X_{1, t-2} + 0.125X_{1,t-3} + 0.5X_{2,t} + \epsilon_{1,t}$,
\item[6$^{\prime\prime}$.] $X_{1,t} = 0.5X_{1,t-1} + 0.25X_{1, t-2} + 0.125X_{1,t-3} + 0.5X_{2,t}^2 + \epsilon_{1,t}$,
\item[7$^{\prime\prime}$.] $X_{1,t} = 0.5X_{1,t-1}X_{2, t} + 0.25X_{1,t-2} + 0.125X_{1,t-3} + \epsilon_{1,t}$,
\item[8$^{\prime\prime}$.] $X_{1,t} = 0.5X_{1,t-1} + 0.25X_{1,t-2} + 0.125X_{1,t-3} + 0.5X_{2, t}\epsilon_{1,t}$,
\item[9$^{\prime\prime}$.] $X_{1,t} = \sqrt{h_t}\epsilon_{1,t}$, $h_t = 0.01 + 0.5X_{1,t-1}^2 + 0.25X_{1,t-2}^2 + 0.125X_{1,t-3}^2 + 0.25X_{2,t}^2$, and
\item[10$^{\prime\prime}$.] Same as DGP10 and DGP10$^{\prime}$ above, except for the new definition of $\X_{3,t}$.
\end{itemize}

Again, we observe simulation results in Table \ref{tab:dim5}. Contrary
to Su and White (\protect\hyperlink{ref-su2014testing}{2014}), we have
also run our test for \(n=100\) which reveals that we can obtain some
power also in that case. All in all, the simplified test based on the
LGPC performs mostly on par with other non-parametric tests in this
setting as well.

The full potential of the LGPC-test may not have been reached in these
experiments. If one compares with independence testing using the local
Gaussian correlation a considerable increase in power was obtained by
focusing the tests appropriately, see Berentsen and Tjøstheim
(\protect\hyperlink{ref-berentsen2014recognizing}{2014}). This was done
by exploiting the typical local dependence pattern for financial
variables. To do this here, one must ascertain whether such patterns
exist for local conditional dependence. Finally, the simplified test
based on pairwise relations may be better able to fight the curse of
dimensionality as the dimension \(p\) of \(\Xtwo = (X_3, \ldots, X_p)\)
increases, cf. Otneim and Tjøstheim
(\protect\hyperlink{ref-otneim2017conditional}{2018}).

\section{Conclusion and outlook}\label{conclusion-and-outlook}

The purpose of this paper is two-fold. First we define and develop the
LGPC, a local measure of conditional dependence that has useful
properties and is easy to interpret. We then explore the possibility of
using the LGPC in estimation of conditional dependence and in testing
for conditional independence, which contributes to the recent
econometric literature on this topic. The results are promising, and
suggest that the newly developed test provide powerful improvements to
existing methods.

For 3 scalar variables \(X_1, X_2, X_3\) a full trivariate analysis
depending on all 3 coordinates can be undertaken as in Section
\ref{chap:trivariate-full}. For a higher dimensional conditioning
variable \(\X_3\) a pairwise approximation and simplification has been
described in Section \ref{chap:bivariate}. This can be likened to
additive approximation in nonparametric regression.

Both the full trivariate approach and the pairwise simplification have
been mainly analyzed at the \(z\)-scale using the transformed variables
\(\Z = \Phi^{-1}(F(\X))\). The corresponding LGPC \(\alpha(\z)\) is
invariant to monotone transformations to the marginals. The LGPC
\(\alpha(\x)\) on the \(x\)-scale can be obtained from \(\alpha(\z)\) by
taking \(\x = F^{-1}(\Phi(\z))\), and it is in general different from
\(\alpha(\z)\) and is not invariant to monotone transformations of the
marginals. An analogue is the different transformation properties of the
Pearson correlation and the copula structure in describing joint
dependence. For multivariate Gaussian variables,
\(\alpha(\x) \equiv \alpha(\z) \equiv \alpha\), which is the ordinary
global partial correlation.

There is a potential for much further work. The \(X_1\) and \(X_2\)
variables can be made into vectors leading to a local dependence between
groups of variables. A next natural step may be structural equations as
used in network analysis and Pearls-type causality (Pearl
\protect\hyperlink{ref-pearl2000causality}{2000}), and ultimately
further exploring the relationship between that type of causality and
Granger causality.

All methods presented in this paper have been implemented in the R
programming language, and is available in the package \textbf{lg} (Håkon
Otneim \protect\hyperlink{ref-otneim2019lg}{2019}). In addition, we
provide code and data for easy replication of the results presented here
in an online appendix.

\section{Appendix}\label{appendix}

\subsection{Some details regarding the proof of Theorem
\ref{thm:loccor3}}\label{app-loccor3}

For fixed bandwidths we have from standard arguments, such as those
provided by Hjort and Jones
(\protect\hyperlink{ref-hjort1996locally}{1996}), that

\begin{equation}
\sqrt{nb^3}\left(\hfrho_n - \frho_0\right) \stackrel{\mathcal{L}}{\rightarrow} \mathcal{N}(0, \Jb^{-1}\Mb(\Jb^{-1})^T),
\label{eq:fixed-h}
\end{equation}

which corresponds to the usual rate for nonparametric density
estimation. As \(\hh\rightarrow0\), however, we must take extra care
when considering the asymptotic covariance matrix. Tjøstheim and
Hufthammer (\protect\hyperlink{ref-tjostheim2013local}{2013}), Section
4, write the Taylor expansions of \(\Mb\) and \(\Jb\), being functionals
of three variables \((x_1, x_2, x_3)\) or \((z_1,z_2,z_3)\), as
\[\Mb = I_{\bm{M}} + II_{\bm{M}} + o(b^2) \,\,\, \textrm{ and } \Jb = I_{\bm{J}} + II_{\bm{J}} + o(b^2),\]
and consider each term. For example, we write their first term in the
expansion of \(\Mb\) in the three-variate case as follows:
\[I_{\bm{M}} \sim \int K^2(w_1, w_2, w_3)Ab_wb_w^TA^Tf(\z + b_1w_1 + b_2w_2 + b_3w_3)\,\textrm{d}w_1\,\textrm{d}w_2\,\textrm{d}w_3,\]
where \(K\) is a product kernel, and, in our case, \(b_w\) is the vector
defined by
\(b_w^T = \begin{pmatrix} 1 & b_1w_1 & b_2w_2 & b_3w_3 \end{pmatrix}\),
and \(A\) is the \(3\times 4\) matrix
\(A = \begin{pmatrix} v & v_{z_1} & v_{z_2} & v_{z_3}\end{pmatrix}\)
with \(\fv^T = \begin{pmatrix} v_1 & v_2 & v_3 \end{pmatrix}\),
\(v_i = \partial L_n/\partial \rho_i\), where the index \(i=1,2,3\)
represents the three local correlations, and
\(v_{z_i} = \partial v/\partial z_i\). In the next step, they compute
the matrix
\(\int K^2(w_1, w_2, w_3)b_wb_w^T\,\textrm{d}w_1\,\textrm{d}w_2\,\textrm{d}w_3\),
which, by omitting all constant factors and exploiting that
\(\int w^k K^2(w)\,\textrm{d}w = 0\) for \(k = 1,3\), in our case
becomes the diagonal matrix
\[H = \begin{pmatrix} 1 & 0 & 0 & 0 \\ 0 & b_1^2w_1^2 & 0 & 0 \\ 0 & 0 & b_2^2w_2^2 & 0 \\ 0 & 0 & 0 & b_3^2w_3^2 \end{pmatrix}.\]

The second term defining \(\Mb\) has smaller order than \(I_{\bm{M}}\)
as \(b \rightarrow 0\), while the first term of \(\Jb\) can be treated
exactly as \(I_{\bm{M}}\) above resulting in a matrix of order \(b^2\).

The second term in \(\Jb\) can similarly be written as
\(II_J = B(\psi(\z, \theta_h(\z)) - f(\z)) + o(b^2)\), where
\(B = \nabla \fv(\z, \theta_b(\z))\) are second derivatives of the local
likelihood function with respect to the local correlations, and
\(\psi(\z, \theta_b(\z)) - f(\z) = O(b^2)\) according to Hjort and Jones
(\protect\hyperlink{ref-hjort1996locally}{1996}). Hence, it follows that
the leading term of \(\Jb^{-1}\Mb(\Jb^{-1})^T\) is \(O(b^{-2})\), which
again means that the convergence rate in eq. \eqref{eq:fixed-h} must be
modified accordingly in order to balance the convergence in eq.
\eqref{eq:an3}. Extracting an analytic expression for the leading term of
the asymptotic covariance matrix is, as mentioned by Tjøstheim and
Hufthammer (\protect\hyperlink{ref-tjostheim2013local}{2013}), possible
by means of symbolic manipulation software, but the expression is very
complicated and of little practical use. See also the corresponding
proof of Theorem 3 in Tjøstheim and Hufthammer
(\protect\hyperlink{ref-tjostheim2013local}{2013}).

\subsection{The asymptotic variance of the LGPC}\label{app-asvar}

The limiting distribution of the LGPC is given in eq.
\eqref{eq:limitlgpc}. In this section, we will present the basic steps
needed to work out the value of \(\nabla g(\frho)\). For the sake of
this particular argument, we may simplify notation quite a bit, leaving
out what we do not need: We drop the \(\z\)-dependence, and simply write
\(\fSigma = \fSigma_{11.2}(\z)\) for the \(2\times2\) local partial
covariance matrix between \(Z_1\) and \(Z_2\) given
\((Z_3,\ldots,Z_p)\). The vector of local correlations \(\frho\) is
indexed by \(k\), which in turn means that \(k\) corresponds to a
specific \emph{pair} \((i,j)\): \(\rho_k = \rho_{ij}\), with
\(\rho_{ij} = \rho_{ji}\). Let \(\fSigma^{(k)}\) be the element-wise
matrix of partial derivatives
\[\fSigma^{(k)} = \frac{\partial \fSigma}{\partial \rho_k} = \frac{\partial \fSigma}{\partial \rho_{ij}},\]
which means that element \(k\) in the gradient \(\nabla g(\frho)\) is
given by

\begin{equation}
\left\{\nabla g(\frho) \right\}_k = \frac{\fSigma_{11}\fSigma_{22}\fSigma^{(k)}_{12} - \fSigma_{12}\left(\fSigma_{11}^{(k)}\fSigma_{22} + \fSigma_{11}\fSigma_{22}^{(k)}\right)}{\left(\fSigma_{11}\fSigma_{22}\right)^2},
\label{eq:gradientelement}
\end{equation}

where the double subscript to \(\fSigma\) and \(\fSigma^{(k)}\) here
means matrix elements. In order to avoid confusion with the double
subscripts defining matrix \emph{partitions} in
\eqref{eq:matrixlocalpartial}, we re-label the matrix blocks \(\R_{11}\) ,
\(\R_{12}\) and \(\R_{22}\) to \(\R_1\), \(\R_2\) and \(\R_3\)
respectively, with \(\R_{21} = \R_2^T\). Following basic differentiation
rules for matrices (see e.g. Van den Bos
(\protect\hyperlink{ref-van2007parameter}{2007})), we have that,
element-wise,

\begin{equation}
\frac{\partial\left(\R_2\R_3^{-1}\R_2^T\right)}{\partial\rho_k} = \frac{\partial\R_2}{\partial\rho_k}\R_3^{-1}\R_3^T - \R_2\R_3^{-1}\frac{\partial \R_3}{\partial \rho_k}\R_3^{-1}\R_2^T + \R_2\R_3^{-1}\frac{\partial\R_2^T}{\partial\rho_k}.
\label{eq:derivative}
\end{equation}

The value of the gradient \eqref{eq:gradientelement} depends on \(k\), and
when differentiating \eqref{eq:matrixlocalpartial} with respect to
\(\rho_k\) there are three different cases we must consider:

\begin{enumerate}
\def\labelenumi{\arabic{enumi}.}
\tightlist
\item
  If \(\rho_k = \rho_{12}\) the second term in
  \eqref{eq:matrixlocalpartial} vanishes, because \(\rho_{12}\) is only
  present in \(\R_1\). Indeed,
  \[\R_{11} = \begin{pmatrix}1 & \rho_{12} \\ \rho_{12} & 1 \end{pmatrix},\]
  which means that
  \[\fSigma^{(k)} = \begin{pmatrix}0&1\\1&0\end{pmatrix} - \begin{pmatrix}0&0\\0&0\end{pmatrix},\]
  and we easily see from \eqref{eq:gradientelement} that in this
  particular case, \(\left\{\nabla g(\frho) \right\}_k = 1\).
\item
  If \(\rho_k = \rho_{ij}\) where \(i \in \{1,2\}\) and
  \(j \in \{3, \ldots,p\}\), then \(\rho_{ij}\) is an element in
  \(\R_2\), but not \(\R_1\) or \(\R_3\). For that reason, we are left
  with the second term in \eqref{eq:matrixlocalpartial}, and furthermore,
  the middle term in the derivative \eqref{eq:derivative} above will also
  vanish. The \(2\times(p-2)\) matrix
  \(\partial\R_{12}/\partial\rho_{ij}\) will contain only zeros except
  for a \(1\) in the position of \(\rho_{ij}\); and
  \(\partial\R_{21}/\partial\rho_{ij}\) is its transposed. Write
  \(\fC = \R_{22}^{-1}\R_{21}\), and we see that \(\fSigma^{(k)}\) in
  this case is a simple linear combination of the elements in \(\fC\):
  \[\fSigma^{(k)} = \frac{\partial\R_{12}}{\partial\rho_{ij}}\fC + \fC^T\frac{\partial\R_{21}}{\partial\rho_{ij}},\]
  from which we select the elements needed to calculate
  \(\left\{\nabla g(\frho) \right\}_k\) through
  \eqref{eq:gradientelement}.\\
\item
  In this case, we look at the derivative of \(g(\frho)\) with respect
  to the local correlations located in \(\R_3\), that is, \(\rho_{ij}\)
  for \(i,j \in \{3,\ldots,p\}\). We only need to keep the middle term
  in \eqref{eq:derivative}. We have that
  \[\fSigma^{(k)} = -\fC^T\frac{\partial\R_3}{\partial\rho_j}\fC,\]
  where \(\partial\R_3/\partial\rho_k\) is a symmetric
  \((p-2)\times(p-2)\) matrix of zeros everywhere, except for 1s in the
  positions corresponding to \(\rho_k\) with respect to which we perform
  the differentiation. Finally, equation \eqref{eq:derivative} provides
  the final expression for \(\left\{\nabla g(\frho) \right\}_k\).
\end{enumerate}

\subsection{Proof of Theorem \ref{thm:pseudo}}\label{proofpseudo}

Theorem \ref{thm:pseudo} states that using marginally standard normal
pseudo-observations instead of exactly standard normally distributed
observations for estimating the LGPC does not change the conclusions in
Theorems \ref{thm:loccor3} and \ref{thm:loccor}. The result may be
expected, as the empirical distribution functions \(\widehat{F}_i\),
\(i = 1,\ldots,p\) converge at a rate faster than the local correlations
(\(n^{-\frac{1}{2}}\) vs. \((nb^5)^{-\frac{1}{2}}\) or
\((nb^2)^{-\frac{1}{2}}\)) depending on whether the rate in Theorem
\ref{thm:loccor3} or \ref{thm:loccor} is used), but we must attend to
some details nevertheless.

Proving joint asymptotic normality of the local Gaussian correlations
with marginally normally distributed variables
\((Z_1, \ldots, Z_p) = \left(\Phi^{-1}(F_1(X_1)), \ldots, \Phi^{-1}(F_p(X_p))\right)\),
where \(F_i\) is the cdf of \(X_i\) for \(i = 1,\ldots,p\), relies on
proving asymptotic normality for the variables
\[Y_n(\z) = \frac{1}{n}\sum_{i=1}^nK_b(\Z_i - \z)\fv(\Z_i, \frho_0),\]
where \(\fv(\cdot) = \partial L(\cdot)/\partial\frho\) in the same way
as in Section \ref{app-loccor3}. This has been done by Otneim and
Tjøstheim (\protect\hyperlink{ref-otneim2017locally}{2017}) in the iid
case using arguments from Schervish
(\protect\hyperlink{ref-schervish1995theory}{2012}), and by Otneim and
Tjøstheim (\protect\hyperlink{ref-otneim2017conditional}{2018}) in the
\(\alpha\)-mixing case, using arguments from Fan and Yao
(\protect\hyperlink{ref-fan2008nonlinear}{2008}). We need here to prove
asymptotic normality of the variables

\[\hY_n(\z) = \frac{1}{n}\sum_{i=1}^nK_b(\hZ_i - \z)\fv(\hZ_i,\frho_0),\]
with \(\hnZ_i = \Phi^{-1}(F_n(\X_i))\), and \(F_n(\X_i)\) is the
application of the marginal empirical distribution function to the
corresponding components of the vector \(\X_i\).

In order to ease notation we state the proof of Theorem \ref{thm:pseudo}
in the simplified case only, where the local correlations are estimated
using \emph{pairs} of variables, which is the case that is treated in
Theorem \ref{thm:loccor}. It can be extended to the trivariate case
(that we treat in Theorem \ref{thm:loccor3}) by adding the appropriate
terms in the derivations below, without changing the final conclusion.
Write \(\hY_n(\z) = Y_n(\z) - (Y_n(\z) - \hY_n(\z))\) and do a Taylor
expansion of \(Y_n(\z)\) around \(F_n(\X_i)\). Since we use product
kernels, we then need derivatives of (assuming without loss of
generality that \(K(x) = K(-x)\))

\[H_{\z,\hh} = K\left(\frac{z_1 - \Phi^{-1}(y_1)}{b}\right)K\left(\frac{z_2 - \Phi^{-1}(y_2)}{b}\right)v(\Phi^{-1}(\y)),\]
where \(\y = F(\x)\), \(y_i = F(x_i)\), and we write
\(K(x_1,x_2) = K(x_1)K(x_2)\). Writing \(k(z) = K'(z)\) and
\(v_{y_i} = \partial v_i/\partial y_i,\)

\begin{align*}
\frac{\partial H_{\bm{z},b}}{\partial y_1} 
&= k\left(\frac{z_1 - \Phi^{-1}(y_1)}{b}\right) K\left(\frac{z_2 - \Phi^{-1}(y_2)}{b}\right) \frac{v(\Phi^{-1}(\y))}{b\phi(\Phi^{-1}(y_1))} \\
& \qquad\qquad\qquad\qquad\qquad\qquad\qquad\qquad +  K\left(\frac{z_1 - \Phi^{-1}(y_1)}{b}\right) K\left(\frac{z_2 - \Phi^{-1}(y_2)}{b}\right) \frac{v_1(\Phi^{-1}(\y))}{\phi(\Phi^{-1}(y_1))}, \\
\frac{\partial H_{\bm{z},b}}{\partial y_2} 
&= K\left(\frac{z_1 - \Phi^{-1}(y_1)}{b}\right) k\left(\frac{z_2 - \Phi^{-1}(y_2)}{b}\right) \frac{v(\Phi^{-1}(\y))}{b\phi(\Phi^{-1}(y_2))} \\
& \qquad\qquad\qquad\qquad\qquad\qquad\qquad\qquad+  K\left(\frac{z_1 - \Phi^{-1}(y_1)}{b}\right) K\left(\frac{z_2 - \Phi^{-1}(y_2)}{b}\right) \frac{v_2(\Phi^{-1}(\y))}{\phi(\Phi^{-1}(y_1))}.
\end{align*}

A typical term in the Taylor expansion of \(Y_n(\z)\) takes the
following form:

\begin{align*}
& K_b\left(\Phi^{-1}(F(\X_i)) - \z\right)v\left(\Phi^{-1}(F(\X_i)), \rho_0\right) = \\
& \qquad K_b\left(\Phi^{-1}(F_n(\X_i)) - \z\right)v\left(\Phi^{-1}(F_n(\X_i)), \rho_0\right) \\
& \qquad\qquad + k\left(\frac{z_1 - \Phi^{-1}(F_n^*(X_{i,1}))}{b}\right) K\left(\frac{z_2 - \Phi^{-1}(F_n^*(x_{i, 2}))}{b}\right) \frac{v(\Phi^{-1}(F_n^*(\X_i)))}{b\phi(\Phi^{-1}(F_n^*(X_{i,1})))}\left(F(X_{i,1}) - F_n(X_{i, 1})\right) \\
& \qquad\qquad + K\left(\frac{z_1 - \Phi^{-1}(F_n^*(X_{i,1}))}{b}\right) K\left(\frac{z_2 - \Phi^{-1}(F_n^*(x_{i, 2}))}{b}\right) \frac{v_1(\Phi^{-1}(F_n^*(\bm{X}_i)))}{\phi(\Phi^{-1}(F_n^*(X_{i,1})))}\left(F(X_{i,1}) - F_n(X_{i, 1})\right) \\
&\qquad + \textrm{ analogous terms for the second (and possibly third) index involving } \left(F(X_{i,2}) - F_n(X_{i,2})\right), 
\end{align*}

where \(F_n^*\) comes from the mean value theorem. The challenge in
proving the desired result is to control the behavior of quantities on
the form \(1/\phi(\Phi^{-1}(F(x)))\), because this fraction tends to
infinity as \(|x| \rightarrow \infty\). The key to this problem is the
assumption that the kernel function \(K(\cdot)\), and thus its
derivative \(k\), have compact support. This means, that in the
expressions above, \(K = k \equiv 0\) if for some \(M > 0\)
\[\frac{|z_1 - \Phi^{-1}(F_n^*(X_{i,1}))|}{b} \geq M,\] or
\[|z_1 - \Phi^{-1}(F_n^*(X_{i,1}))| \geq Mb,\] or, by removing the
absolute value signs,
\[ \Phi^{-1}(F_n^*(X_{i,1})) \leq z_1 - Mb\,\, \textrm{ or }\,\, \Phi^{-1}(F_n^*(X_{i,1})) \geq z_1 + Mb,\]
where \(M\) may be large. The same reasoning applies of course to the
second index as well. Letting \(n\rightarrow\infty\) and using
consistency of the empirical distribution function, it follows that the
kernel function term is zero if

\[X_{i,1} \leq F^{-1}(\Phi(z_1 - Mb)) \,\,\, \textrm{ or } \,\,\, X_{i,1} \geq F^{-1}(\Phi(z_1 + Mb)).\]
The same reasoning applies to the derivative of the kernel function,
where we consider the function \(k/b\) instead of \(K\). All of this
implies that the challenge of controlling the magnitude of the Taylor
terms above as \(x\rightarrow\infty\) disappears, since, taking
boundedness of the other terms into account,

\[ k\left(\frac{z_1 - \Phi^{-1}(F_n^*(X_{i,1}))}{b}\right) K\left(\frac{z_2 - \Phi^{-1}(F_n^*(x_{i, 2}))}{b}\right) \frac{v(\Phi^{-1}(F_n^*(\X_i)))}{b\phi(\Phi^{-1}(F_n^*(X_{i,1})))}\]
and
\[K\left(\frac{z_1 - \Phi^{-1}(F_n^*(X_{i,1}))}{b}\right) K\left(\frac{z_2 - \Phi^{-1}(F_n^*(x_{i, 2}))}{b}\right) \frac{v_1(\Phi^{-1}(F_n^*(\X_i)))}{\phi(\Phi^{-1}(F_n^*(X_{i,1})))}\]
are bounded almost surely as \(n\rightarrow\infty\) and
\(b\rightarrow 0\).

All this means that if we can prove that
\[\frac{1}{n}\sum_{i=1}^n(F_n(X_i) - F(X_i)) \stackrel{p}{\rightarrow} 0\]
at an appropriate rate, then we are done by Slutsky's Theorem.

We need a bound on the supremum of \(|F_n(X_i) - F(X_i)|\) and
\(|F_n(x) - F(x)|\) as \(n \rightarrow \infty\). According to Arcones
(\protect\hyperlink{ref-arcones1995law}{1995}), the best law of the
iterated logarithm has been obtained by Rio
(\protect\hyperlink{ref-rio1995functional}{1995}). One statement of this
is as follows:

Let \(f(x) \in L_p\), \(2<p\leq\infty\),
\(\sum\alpha_k k^{2/(p-2)} < \infty\), then the sequence \(f(X_i)\)
satisfies the compact LIL (\emph{law of the iterated logarithms}), i.e.
\[\left\{\left(n2\log\log n\right)^{-\frac{1}{2}} \sum_{i=1}^n(f(X_i) - E[f(X_i)]) \right\}_{n=1}^{\infty}\]
is relatively compact and its limit set is \([-\sigma, \sigma]\) where
\[\sigma^2 = \textrm{Var}\Big(f(X_i)\Big) + 2\sum_{j=1}^{\infty}\textrm{Cov}\Big(f(X_i), f(X_j)\Big).\]
For a fixed \(x\), take \(f(X_i) = I(X_i\leq x)\). Then
E\((f(X_i)) = F(x)\) and \(f(x)\in L_p\) for all \(p\), and
\[\left\{\left(n2\log\log n\right)^{-\frac{1}{2}} \sum_{j=1}^n \left(I(X_j \leq x) - F(x) \right) \right\}\]
has its limit set in \([-\sigma, \sigma]\). Here
\[\textrm{Var}(I(X_i \leq x)) = F(x)(1-F(x)) \leq \frac{1}{4},\] and
according to Davydov's Lemma (see e.g. Fan and Yao
(\protect\hyperlink{ref-fan2008nonlinear}{2008}) p.~278),
\[\big| \textrm{Cov}(I(X_i \leq x), I(X_j \leq x))\big| \leq 8 \|I(X_i \leq x)\|_p \|I(X_j \leq x)\|_q \alpha(|j-i|)^{1-p^{-1}-q^{-1}}, \,\,\, p,q>1, \,\,\, \frac{1}{p}+\frac{1}{q} < 1,\]
with
\(\| I(X_i \leq x)\|_p = \left(\textrm{E}|I(X_i \leq x)|^p\right)^{1/p}\).
We may for example take \(p=q=4\), leading to
\(\| I(X_i \leq x) \|_4 = F(x)^{1/4} \leq 1\).

Since we have assumed exponential mixing (much weaker conditions
suffice), then Rio's condition \(\sum \alpha_k k^{2/(p-2)} < \infty\) is
satisfied, and the bounds of \(-\sigma\) and \(\sigma\) can be made
independent of \(x\).

Having \[F_n(x) = \frac{1}{n}\sum_{j=1}^n I(X_j \leq x)\] means that
\[n^{\frac{1}{2}}[2\log \log n]^{-\frac{1}{2}} (F_n(x) - F(x))\] has its
limits in \([-\sigma, \sigma]\) or that the supremum of
\(|F_n(x) - F(x)|\) is almost surely bounded by
\(n^{-1/2}(2\log\log n)^{1/2}\).

This immediately implies that
\[\frac{1}{n}\sum_{i=1}^n\left(F_n(X_i) - F(X_i)\right)\] is almost
surely (and consequentially in probability) bounded by
\(n^{-\frac{1}{2}}(2 \log \log n)^{\frac{1}{2}}\) which for all
conceivable bandwidths is faster than the convergence rates of the CLT
for \(Y_n(\z)\), which is \((nb^2)^{-\frac{1}{2}}\).

\section*{References}\label{references}
\addcontentsline{toc}{section}{References}

\hypertarget{refs}{}
\hypertarget{ref-arcones1995law}{}
Arcones, Miguel A. 1995. ``The Law of the Iterated Logarithm for
Empirical Processes Under Absolute Regularity.'' \emph{Journal of
Theoretical Probability} 8 (2): 433--51.
doi:\href{https://doi.org/10.1007/BF02212887}{10.1007/BF02212887}.

\hypertarget{ref-baba2004partial}{}
Baba, Kunihiro, Ritei Shibata, and Masaaki Sibuya. 2004. ``Partial
Correlation and Conditional Correlation as Measures of Conditional
Independence.'' \emph{Australian \& New Zealand Journal of Statistics}
46 (4). Wiley Online Library: 657--64.

\hypertarget{ref-berentsen2014recognizing}{}
Berentsen, Geir Drage, and Dag Tjøstheim. 2014. ``Recognizing and
Visualizing Departures from Independence in Bivariate Data Using Local
Gaussian Correlation.'' \emph{Statistics and Computing} 24 (5).
Springer: 785--801.

\hypertarget{ref-berentsen2014introducing}{}
Berentsen, Geir Drage, Tore Selland Kleppe, and Dag Tjøstheim. 2014.
``Introducing Localgauss, an R Package for Estimating and Visualizing
Local Gaussian Correlation.'' \emph{Journal of Statistical Software} 56
(1). Foundation for Open Access Statistics: 1--18.

\hypertarget{ref-bergsma2011nonparametric}{}
Bergsma, Wicher. 2010. ``Nonparametric Testing of Conditional
Independence by Means of the Partial Copula.'' \emph{Unpublished
Manuscript}. \url{http://dx.doi.org/10.2139/ssrn.1702981}.

\hypertarget{ref-boue:taam:2014}{}
Bouezmarni, Taoufik, and Abderrahim Taamouti. 2014. ``Nonparametric
Tests for Conditional Independence Using Conditional Distributions.''
\emph{Journal of Nonparametric Statistics} 26 (4). Taylor \& Francis:
697--719.

\hypertarget{ref-bouezmarni2012nonparametric}{}
Bouezmarni, Taoufik, Jeroen V K Rombouts, and Abderrahim Taamouti. 2012.
``Nonparametric Copula-Based Test for Conditional Independence with
Applications to Granger Causality.'' \emph{Journal of Business and
Economic Statistics} 30 (2): 275--87.

\hypertarget{ref-cheng2012conditional}{}
Cheng, Yu-Hsiang, and Tzee-Ming Huang. 2012. ``A Conditional
Independence Test for Dependent Data Based on Maximal Conditional
Correlation.'' \emph{Journal of Multivariate Analysis} 107. Elsevier:
210--26.

\hypertarget{ref-delgado2001significance}{}
Delgado, Miguel A, and Wenceslao González Manteiga. 2001. ``Significance
Testing in Nonparametric Regression Based on the Bootstrap.'' \emph{The
Annals of Statistics} 29 (5). Institute of Mathematical Statistics:
1469--1507.

\hypertarget{ref-fan2008nonlinear}{}
Fan, Jianqing, and Qiwei Yao. 2008. \emph{Nonlinear Time Series:
Nonparametric and Parametric Methods}. Springer Science \& Business
Media.

\hypertarget{ref-granger1969investigating}{}
Granger, Clive WJ. 1969. ``Investigating Causal Relations by Econometric
Models and Cross-Spectral Methods.'' \emph{Econometrica}. JSTOR,
424--38.

\hypertarget{ref-granger1980testing}{}
---------. 1980. ``Testing for Causality: A Personal Viewpoint.''
\emph{Journal of Economic Dynamics and Control} 2. Elsevier: 329--52.

\hypertarget{ref-hjort1996locally}{}
Hjort, Nils Lid, and MC Jones. 1996. ``Locally Parametric Nonparametric
Density Estimation.'' \emph{The Annals of Statistics}. JSTOR, 1619--47.

\hypertarget{ref-huang2010testing}{}
Huang, Tzee-Ming. 2010. ``Testing Conditional Independence Using Maximal
Nonlinear Conditional Correlation.'' \emph{The Annals of Statistics} 38
(4). Institute of Mathematical Statistics: 2047--91.

\hypertarget{ref-jordanger2017nonlinear}{}
Jordanger, Lars Arne, and Dag Tjøstheim. 2017. ``Nonlinear Spectral
Analysis via the Local Gaussian Correlation.'' \emph{arXiv Preprint
arXiv:1708.02166, Under revision for the Journal of the American
Statistical Association}.

\hypertarget{ref-lacal2018estimating}{}
Lacal, Virginia, and Dag Tjøstheim. 2018. ``Estimating and Testing
Nonlinear Local Dependence Between Two Time Series.'' \emph{Journal of
Business \& Economic Statistics}. Taylor \& Francis, 1--13.

\hypertarget{ref-lawrance1976conditional}{}
Lawrance, AJ. 1976. ``On Conditional and Partial Correlation.''
\emph{The American Statistician} 30 (3). Taylor \& Francis: 146--49.

\hypertarget{ref-linton1997conditional}{}
Linton, Oliver, and Pedro Gozalo. 1997. ``Conditional Independence
Restrictions: Testing and Estimation.'' Unpublished manuscript.

\hypertarget{ref-linton2014testing}{}
---------. 2014. ``Testing Conditional Independence Restrictions.''
\emph{Econometric Reviews} 33 (5-6). Taylor \& Francis: 523--52.

\hypertarget{ref-otneim2016multivariate}{}
Otneim, Håkon. 2016. ``Multivariate and Conditional Density Estimation
Using Local Gaussian Approximations.'' PhD thesis, University of Bergen;
The University of Bergen.

\hypertarget{ref-otneim2019lg}{}
Otneim, Håkon. 2019. \emph{Lg: Locally Gaussian Distributions:
Estimation and Methods}.

\hypertarget{ref-otneim2017locally}{}
Otneim, Håkon, and Dag Tjøstheim. 2017. ``The Locally Gaussian Density
Estimator for Multivariate Data.'' \emph{Statistics and Computing} 27
(6). Springer: 1595--1616.

\hypertarget{ref-otneim2017conditional}{}
---------. 2018. ``Conditional Density Estimation Using the Local
Gaussian Correlation.'' \emph{Statistics and Computing} 28 (2).
Springer: 303--21.

\hypertarget{ref-patra2016on}{}
Patra, Rohit K, Bodhisattva Sen, and Gábor J Székely. 2016. ``On a
Nonparametric Notion of Residual and Its Applications.''
\emph{Statistics and Probability Letters} 109: 208--13.

\hypertarget{ref-pearl2000causality}{}
Pearl, Judea. 2000. \emph{Causality: Models, Reasoning and Inference}.
Vol. 29. Springer.

\hypertarget{ref-rio1995functional}{}
Rio, Emmanuel. 1995. ``The Functional Law of the Iterated Logarithm for
Stationary Strongly Mixing Sequences.'' \emph{The Annals of
Probability}. JSTOR, 1188--1203.

\hypertarget{ref-schervish1995theory}{}
Schervish, Mark J. 2012. \emph{Theory of Statistics}. Springer Science
\& Business Media.

\hypertarget{ref-song2009testing}{}
Song, Kyungchul. 2009. ``Testing Conditional Independence via Rosenblatt
Transforms.'' \emph{The Annals of Statistics} 37 (6B). Institute of
Mathematical Statistics: 4011--45.

\hypertarget{ref-su2007consistent}{}
Su, Liangjun, and Halbert White. 2007. ``A Consistent Characteristic
Function-Based Test for Conditional Independence.'' \emph{Journal of
Econometrics} 141 (2). Elsevier: 807--34.

\hypertarget{ref-su2008nonparametric}{}
---------. 2008. ``A Nonparametric Hellinger Metric Test for Conditional
Independence.'' \emph{Econometric Theory} 24 (4). Cambridge University
Press: 829--64.

\hypertarget{ref-su2014testing}{}
---------. 2014. ``Testing Conditional Independence via Empirical
Likelihood.'' \emph{Journal of Econometrics} 182 (1). Elsevier: 27--44.

\hypertarget{ref-su2012conditional}{}
Su, Liangjun, and Halbert L White. 2012. ``Conditional Independence
Specification Testing for Dependent Processes with Local Polynomial
Quantile Regression.'' In \emph{Essays in Honor of Jerry Hausman},
355--434. Emerald Group Publishing Limited.

\hypertarget{ref-terasvirta2010modelling}{}
Teräsvirta, Timo, Dag Tjøstheim, and Clive W.J. Granger. 2010.
\emph{Modelling Nonlinear Economic Time Series}. Oxford University
Press.

\hypertarget{ref-tjostheim2013local}{}
Tjøstheim, Dag, and Karl Ove Hufthammer. 2013. ``Local Gaussian
Correlation: A New Measure of Dependence.'' \emph{Journal of
Econometrics} 172 (1). Elsevier: 33--48.

\hypertarget{ref-van2007parameter}{}
Van den Bos, Adriaan. 2007. \emph{Parameter Estimation for Scientists
and Engineers}. John Wiley \& Sons.

\hypertarget{ref-wang2017characteristic}{}
Wang, Xia, and Yongmiao Hong. 2017. ``Characteristic Function Based
Testing for Conditional Independence: A Nonparametric Regression
Approach.'' \emph{Econometric Theory}. Cambridge University Press,
1--35.

\hypertarget{ref-wang2015conditional}{}
Wang, Xueqin, Wenliang Pan, Wenhao Hu, Yuan Tian, and Heping Zhang.
2015. ``Conditional Distance Correlation.'' \emph{Journal of the
American Statistical Association} 110 (512). Taylor \& Francis:
1726--34.

\hypertarget{ref-yahoo}{}
Yahoo Finance. 2018. \emph{Accessed June 2018}.

\hypertarget{ref-zhang2012kernel}{}
Zhang, Kun, Jonas Peters, Dominik Janzing, and Bernhard Schölkopf. 2012.
``Kernel-Based Conditional Independence Test and Application in Causal
Discovery.'' \emph{arXiv Preprint arXiv:1202.3775}.

\end{document}